\documentclass[letterpaper]{IEEEtran}         
\usepackage{amsmath,graphicx}
\usepackage{amsfonts}
\usepackage{enumerate}
\usepackage{bm}
\usepackage{algorithm} 
\usepackage{algorithmic} 

\usepackage{mathrsfs}
\usepackage{stfloats}
\usepackage{enumerate}
\usepackage{epstopdf}
\usepackage{subfigure}
\usepackage{float}
\usepackage{amsbsy}
\usepackage{amsmath}
\usepackage{amssymb}
\usepackage{color}
\usepackage{ntheorem}
\usepackage{cite}

\theorembodyfont{\upshape}
\theoremheaderfont{\rmfamily\itshape}
\theoremseparator{:}
\theoremstyle{remark}

\newtheorem{theo}{\hspace{1em}Theorem}

\newtheorem{corollary}{\hspace{1em}Corollary}
\newtheorem*{pproof}{\hspace{2em}Proof}
\begin{document}
\IEEEoverridecommandlockouts
\title{\huge{Multi-User Modular XL-MIMO Communications: Near-Field Beam Focusing Pattern and User Grouping}\vspace{-1pt}}
\author{\IEEEauthorblockN{Xinrui~Li, Zhenjun~Dong, Yong Zeng,  \emph{Senior Member, IEEE},\\
 Shi Jin, \emph{Senior Member, IEEE}, and Rui Zhang, \emph{Fellow, IEEE}
\thanks{This work was supported by the National Key R$\&$D Program of China with Grant number 2019YFB1803400, and
the Fundamental Research Funds for the Central Universities of China under grant number 2242022k60004.
Part of this work has been accepted by IEEE Globecom, 2023 \cite{Lii2023}.}
\thanks{X. Li, Z. Dong, Y. Zeng, and S. Jin are with the National Mobile Communications Research
Laboratory and Frontiers Science Center for Mobile Information Communication and Security, Southeast University, Nanjing 210096, China. Y. Zeng is also
with the Purple Mountain Laboratories, Nanjing 211111, China (e-mail: \{230218659, zhenjun\_dong, yong\_zeng, jinshi\}@seu.edu.cn). (\emph{Corresponding author: Yong Zeng}.)}
\thanks{R. Zhang is with School of Science and Engineering, Shenzhen Research Institute of Big Data, The Chinese University of Hong Kong, Shenzhen, Guangdong 518172, China (e-mail: rzhang@cuhk.edu.cn). He is also with the Department of Electrical and Computer Engineering, National University of Singapore, Singapore 117583 (e-mail: elezhang@nus.edu.sg).}
}}
\maketitle
\vspace{-1cm}

\begin{abstract}
In this paper, we investigate multi-user modular extremely large-scale multiple-input multiple-output (XL-MIMO) communication systems, where modular extremely large-scale uniform linear array (XL-ULA) is deployed at the base station (BS) to serve multiple single-antenna users. By exploiting the unique modular array architecture and considering the potential near-field propagation, we develop sub-array based uniform spherical wave (USW) models for distinct versus common angles of arrival/departure (AoAs/AoDs) with respect to different sub-arrays/modules, respectively. Under such USW models, we analyze the beam focusing patterns at the near-field observation location by using near-field beamforming. The analysis reveals that compared to the conventional XL-MIMO with collocated antenna elements, modular XL-MIMO can provide better spatial resolution by benefiting from its larger array aperture. However, it also incurs undesired grating lobes due to the large inter-module separation. Moreover, it is found that for multi-user modular XL-MIMO
communications, the achievable signal-to-interference-plus-noise ratio (SINR) for users may be degraded by the grating lobes of the beam focusing pattern.
To address this issue, an efficient user grouping method is proposed for multi-user transmission scheduling, so that users located within
the grating lobes of each other are not allocated
to the same time-frequency resource block (RB) for their communications. Numerical results are presented to verify the effectiveness of the
proposed user grouping method, as well as the superior
performance of modular XL-MIMO over its
collocated counterpart with densely distributed users.
\end{abstract}
\section{Introduction}
Massive multiple-input multiple-output (MIMO) is a crucial enabling technology for the
fifth-generation (5G) mobile communication systems to significantly increase the spectral efficiency and spatial resolution \cite{Zhang2020, Hu2021, Bii2019}. In future sixth-generation (6G) networks, the continuous evolution of massive MIMO may render systems involving antennas that are typically termed  extremely large-scale array (XL-array) \cite{Zeng2021, Wang2022}, ultra-massive MIMO \cite{Wang2021, Yuan2022}, extremely large aperture array (ELAA) \cite{Bjornson2019, Liu2021}, or extremely large-scale MIMO (XL-MIMO) \cite{Lu2021, Cui2022}.
Conventionally, there are two typical architectures to accommodate a large number of array elements, i.e., collocated XL-array and distributed antenna system (DAS) \cite{Li2022}. The collocated XL-array typically requires a large contiguous platform for antenna deployment, which may not be available for many practical scenarios. On the other hand, for DAS, its operation requires the sophisticated inter-site signaling coordination and large fronthaul/backhaul capacity \cite{Ngo2017,Ngo20172}.\par

Modular XL-array is a novel promising architecture that may overcome the limitations of collocated and distributed arrays, where array elements are regularly arranged in a modular manner on a common platform, with each module consisting of moderate uniform array antennas, and neighbouring
modules are separated
with relatively large distance \cite{Bertilsson2018, Li2022, Huang2018}. In contrast to the collocated XL-array with the same number of antenna elements, modular XL-array can not only attain the conformal and flexible deployment, but also significantly enhance the spatial resolution, thanks to its larger total array aperture. In particular, modular XL-array can be mounted on the building facades with adjacent modules separated by windows \cite{Li2022}.
Besides, compared to DAS, all antenna elements in modular XL-array share a common site, so that they can more easily realize the joint signal processing without relying on sophisticated inter-site coordination and synchronization \cite{Li2022}.  Note that other similar terms to modular XL-array like large-spaced array \cite{YZeng2022, Ye2022}, radioweaves \cite{Van2019} and irregular antenna array \cite{Ma2019} have also been widely used in the fields of microwave, radar positioning, and signal processing \cite{Peng2018, Li2021, Ge2016}. \par

For modular XL-array with the large array aperture, the conventional uniform planar wave (UPW)-based far-field channel model is no longer sufficient in general \cite{Magoarou2020, Botond2020}. Specifically, the far-field and near-field regions of antenna arrays are typically separated based on the classical Rayleigh distance $r_{\rm rayl}=\frac{2D^2}{\lambda}$, where $D$ is the total array aperture and $\lambda$ is the signal wavelength \cite{Zeng2021,122}. For modular XL-array, the dimension $D$ also includes the relatively large module separations, which may easily make it much larger than that of its collocated counterpart that has the same number of array elements. Therefore, in contrast to the collocated XL-array, modular XL-array is expected to have stronger near-field characteristics, thus rendering the users/scatterers more likely to locate within its near-field region. As such, instead of the conventional UPW model, a more general non-uniform spherical wave (NUSW) model is demanded to accurately characterize the variations of signal amplitude and phase across different array elements \cite{Bj2020, Zeng2021}. For example, the authors in \cite{2022} studied the NUSW-based near-field modelling for collocated extremely large-scale MIMO (XL-MIMO), under which the maximal signal-to-interference-plus-noise ratio (SINR) was derived and the beam correlation coefficient was analyzed. Besides, in multi-user collocated XL-MIMO communication systems, the authors in \cite{2023} formulated the near-field beam focusing problem with the aim of maximizing the users' sum rate, demonstrating the capability of XL-MIMO in mitigating interference from both angle
and distance dimensions. By exploiting the polar-domain sparsity in channel
angular and distance information, the near-field channel parameters of collocated XL-MIMO were estimated in \cite{Dai2022}. Moreover, for multi-path wireless environments, the NUSW-based near-field spatial
correlation was studied in \cite{Dong2022}, where the characteristic of
spatially wide-sense non-stationary process was revealed. However, the aforementioned studies only consider the conventional collocated XL-array, which are not applicable for the emerging modular XL-array.
In our pervious works \cite{Li2022, Li2023}, we considered wireless communications with modular XL-array, where the achievable signal-to-noise ratio (SNR) expression under the near-field NUSW model was derived. However, such existing results only provide the performance characterization for single-user modular XL-array communications. For general multi-user communication systems with modular XL-MIMO, the comprehensive analysis for beam focusing patterns at the near-field observation location with different beamforming techniques is crucial to understand the inter-user interference (IUI) and thereby characterize the ultimate performance of modular XL-MIMO systems, which, to our best knowledge, has not been reported in prior works.\par
To fill the above gaps, in this work, we study multi-user modular XL-MIMO communication systems, where modular
extremely large-scale uniform linear array (XL-ULA) is deployed at the base station (BS) to serve multiple  single-antenna users.
We first develop simplified near-field uniform spherical wave (USW) models for modular XL-ULA by exploiting its unique array architecture, and then analyze the beam focusing patterns at the near-field observation location with near-field beamforming. We show that in contrast to the collocated XL-array having the same number of array elements, modular XL-array is able to greatly enhance the spatial resolution, thanks to its larger total array aperture. However, this also leads to undesired grating lobes due to the large inter-module separation. Fortunately, the grating lobe issue can be effectively mitigated with user grouping for multi-user transmission scheduling, i.e., users located within the grating lobes of each other are not assigned to the same time-frequency resource block (RB) for their communications.  Our main contributions in this paper are outlined as follows:
 \begin{itemize}
\item First, by exploiting the unique geometric architecture of modular XL-array, two methods to simplify the spherical wave model are respectively developed for distinct versus common angles of arrival/departure (AoAs/AoDs) with respect to different sub-arrays/modules. In particular, under the latter model, the near-field array response vector for modular XL-ULA with $N M$ elements can be represented as a Kronecker product of the array response vectors for a sparse ULA with $N$ elements and a collocated ULA with $M$ elements, where $N$ denotes the number of modules and $M$ denotes the number of elements within each module. Such a representation preserves the near-field property while greatly simplifying the subsequent analysis.
\item Second, based on the developed USW models for modular XL-ULA, we analyze the beam focusing patterns at the near-field observation location with the new near-field beamforming. It is revealed that in contrast to its collocated counterpart with the same number of antenna elements, modular XL-ULA can significantly enhance the spatial resolution, while at the cost of more severe grating lobes. Fortunately, different from the conventional UPW-based far-field channel model, the USW-based near-field channel model renders modular XL-ULA exhibiting a higher capability in suppressing grating lobes by benefiting from the accurate characterization for the non-linear phase variations across array modules.

\item Next, for multi-user modular XL-MIMO communications, we derive the user  signal-to-interference-plus-noise ratio (SINR) with three classical beamforming methods based on maximum ratio combining (MRC), zero forcing (ZF) and minimum mean square error (MMSE), respectively. It is shown that all their SINR performance may be degraded by the grating lobes of the beam focusing pattern between each pair of users that are allocated to the same time-frequency RB for transmission. This shows the importance to address the grating lobe issue due to the modular architecture.

\item Last, to resolve the grating lobe issue of multi-user modular XL-MIMO communications, we propose an effective method for users' transmission grouping for achieving their sum rate maximization. As the user grouping optimization problem is combinatorial in nature, which is difficult to be directly solved, we propose an efficient greedy based user grouping method.  Numerical results are presented to demonstrate the effectiveness of the proposed user grouping method, as well as the superior performance of modular XL-MIMO over the conventional collocated XL-MIMO with densely distributed users.
 \end{itemize}\par

It is worth mentioning that grating lobes are well-understood for sparse arrays based on the far-field model \cite{Wang2023}, when the array inter-element distance is greater than half signal wavelength. Research efforts have been devoted to seeking for effective methods to suppress grating lobes. For example, in  \cite{KGanesan2020}, the authors considered the use of radio weaves deployed over four walls in a room, so that
users are better separated as compared to the co-located antenna deployment, which makes up the performance loss caused by grating lobes. In \cite{Kurup2003}, the authors considered a position-phase
synthesis technique to reduce the level of grating lobes in sparse array, which performs better than the methods of only adjusting distance or phase. In \cite{krivosheev2010}, various techniques for suppressing grating lobes are summarized for sparse array, including
relative shifts of adjacent sub-arrays, variable distances between neighboring sub-arrays, different inter-element spacing in
each sub-array, or rotation of the sub-arrays. Additionally, for the radar system with sparse array \cite{Zhuang2008}, a technique of frequency
MIMO where each sub-array owns the independent frequency was proposed for
grating lobe suppression at the cost of wider signal bandwidth. Moreover, in \cite{Gao2022}, to overcome the problem of angular ambiguity caused by grating lobes, the authors proposed a novel algorithm of orthogonal matching pursuit assisted by a collocated array to realize the accurate target detection. However, the prior works mainly considered the grating lobe suppression from the view of optimizing the sparse array only under the far-field UPW model in e.g. radar applications. In contrast, for the unique modular XL-array, all modules are regularly arranged for catering to the actual deployment structure, so that the flexible adjustment of arrays is practically difficult to implement. On the other hand, the grating lobe suppression in the existing far-field UPW model is not suitable for near-field modular XL-MIMO communications studied in this paper. Hence, in multi-user modular XL-MIMO communications, it is necessary to explore the key
characteristics of near-field beam focusing and grating lobes, and resolve the grating lobe issue by new approaches such as exploiting user grouping for multi-user
transmission scheduling, which are not considered in radar systems.  \par

The remainder of this paper is organized as follows. Section II presents our developed
near-field channel models for modular XL-ULA. In Section III, the near-field beam focusing pattern for modular XL-ULA is analyzed.
Section IV derives the SINR expressions and proposes the user grouping algorithm for multi-user modular XL-MIMO communications. Finally, we conclude
this paper in Section V.\par
\emph{Notations}:
$\mathbb{C}^{M\times N}$ is the space of $M\times N$ complex-valued matrices. $||\cdot||$ represents the Euclidean norm. $|\cdot|$ denotes the absolute value of a complex number. $\lfloor \cdot \rfloor$ is the floor operation. $\otimes$ is Kronecker product. $\odot$ is Hadamard product. ${\cal CN}({\boldsymbol 0},{\bf \Sigma})$ is the distribution of a circularly symmetric complex Gaussian (CSCG) random vector with mean $\boldsymbol 0$ and covariance matrix ${\bf \Sigma}$, and ${\cal U}(a,b)$ denotes  the uniform distribution between $a$ and $b$.
\section{Near-Field USW Model for Modular XL-ULA} \label{model}\vspace{-1pt}

As shown in Fig. 1, we consider an uplink wireless communication system with modular XL-ULA deployed at the BS.
The total number of array elements is $N M$, with $N$ being the number of modules and $M$ being the number of antenna elements within each module.
The inter-element spacing for antennas within each module is represented by $d$, which is typically set to half of the signal wavelength, i.e., $d=\frac{\lambda}{2}$. Therefore, the physical size of each module is $S=(M-1)d$. Moreover, let the inter-module distance between the reference elements (say the center elements of different modules) be represented as
$\Gamma d$, where $\Gamma\geq M$ is the modular separation parameter that is dependent on the discontinuous surface of practical installation structure. For example, if modular XL-array is installed on the building facades that are separated by windows, $\Gamma$ is determined by the window size. Thus,
the total physical size
of the modular XL-ULA is $D=[(N-1)\Gamma+(M-1)]d$. In particular, when $\Gamma=M$, the modular XL-array degenerates to the conventional collocated XL-array \cite{Lu2021,Zeng2021}. For convenience of notations, we assume that $N$ and $M$  are odd numbers,
so that the module index $n$ and the antenna index $m$ for each module belong to the integer sets $\mathcal N= \left\{0,\pm 1,\cdots,\pm \frac{N-1}{2}\right\}$ and $\mathcal M=\left\{0, \pm 1,\cdots,\pm \frac{M-1}{2}\right\}$, respectively.
Without loss of generality, the modular XL-ULA is assumed to be placed along the $\emph{y}$-axis, and symmetric around the origin. As a result, the position of the $m$-th element within module $n$ is
${\bf w}_{n,m}=\left[0,y_{n,m}\right]^T$, where $y_{n,m}=(n\Gamma+m)d$, $\forall n\in\mathcal N$, and $\forall m \in \mathcal M$.\par

Suppose that a user or scatterer is located at ${\bf q}=[r\cos\theta, r\sin\theta]^T$, where $\theta \in [-\frac{\pi}{2},\frac{\pi}{2}]$ represents its angle with respect to the positive $x$-axis, and $r$ denotes its distance from the array center.
As such, the distance between $\mathbf q$ and the $m$-th element in module $n$ is
\begin{equation}\label{EQU-1} \vspace{-3pt}
\begin{split}
r_{n,m}&=||{\bf q}-{\bf w}_{n,m}||\\
&=\sqrt{r^2-2r y_{n,m}\sin{\theta}+y_{n,m}^2},\forall n\in\mathcal N, m\in\mathcal M.\\
\end{split}
\end{equation}\par
To accurately characterize the signal amplitude and phase variations across all the $NM$ array elements, the NUSW-based near-field channel model is generally needed. For the modular XL-ULA with a user/scatterer having  distance $r$ and angle $\theta$, the NUSW-based channel vector, denoted as ${\bf h}(r,\theta)\in \mathbb{C}^{(N M)\times 1}$, can be modelled as \cite{Lu2021}
\begin{equation}\label{EQU-2} \vspace{-3pt}
\begin{split}
{\bf h}(r,\theta)&=\left[\frac{\sqrt{\beta_0}}{r_{n,m}}e^{-j\frac{2\pi}{\lambda}r_{n,m}}\right]_{n\in\mathcal N,m\in\mathcal M}=\frac{\sqrt{\beta_0}}{r}{\bf a}(r,\theta),\\
\end{split}
\end{equation}
where $\beta_0$ denotes the reference channel power gain at the distance of $1$ meter ($\rm m$), and ${\bf a}(r,\theta)=\left[\frac{r}{r_{n,m}}e^{-j\frac{2\pi}{\lambda}r_{n,m}}\right]_{n\in\mathcal N,m\in\mathcal M}$ is the array response vector for modular XL-ULA based on the NUSW model \cite{Zeng2021, Li2022}. When $r\ge 1.2D$, it is indicated that the amplitude variations across all array elements can be neglected \cite{122}. In this case, the NUSW model can reduce to the USW model, and the array response vector ${\bf a}(r,\theta)$ simplifies as
\begin{equation}\label{EQU-3-1} \vspace{-3pt}
\begin{split}
&{\bf a}(r,\theta)=\left[e^{-j\frac{2\pi}{\lambda}r_{n,m}}\right]_{n\in\mathcal N,m\in\mathcal M}.\\
\end{split}
\end{equation}\par

\begin{figure}[t]
  \centering
    \includegraphics[scale=0.67]{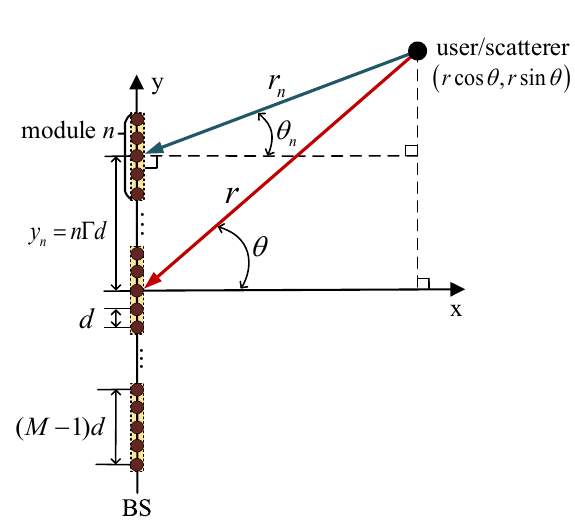}
  \caption{A modular XL-ULA with $N$ modules and $M$ antennas in each module. }\label{pic1}
  \label{12}
  \vspace{-0.3cm}
\end{figure}\par
However, since \eqref{EQU-3-1} is highly dependent on the sophisticated distance expression given by \eqref{EQU-1}, it is still difficult to conduct its performance analysis. To build more tractable near-field array models, we propose two simplifications  of \eqref{EQU-3-1} by using the unique regular structure of modular XL-ULA, which are termed sub-array based USW models for distinct versus common AoAs/AoDs, respectively. As a benchmark, the conventional UPW-based far-field assumption was presented for modelling modular arrays at first.\par
\textit{1) Conventional UPW model}:  When $r\ge\frac{2D^2}{\lambda}$, the user/scatterer is located within the far-field region of the whole array. In this case, the first-order Taylor expansion is used to approximate the distance expression in \eqref{EQU-1}, so that $r_{n,m}\approx r-y_{n,m}\sin\theta $. By substituting such an approximation to \eqref{EQU-3-1}, the UPW-based far-field array response vector for modular ULA can be expressed as
\begin{equation}\label{EQU-4-1} \vspace{-3pt}
\begin{split}
&{\bf a}(\theta)=e^{-j \frac{2\pi}{\lambda}{r}}{\bf p}(\theta)\otimes {\bf b}(\theta),\\
\end{split}
\end{equation}
where ${\bf p}(\theta)=\left[e^{j \frac{2\pi}{\lambda}{n\Gamma d\sin\theta}}\right]_{n\in\mathcal N} \in \mathbb{C}^{N\times 1}$ and ${\bf b}(\theta)=\left[e^{j \frac{2\pi}{\lambda}{md\sin\theta}}\right]_{m\in\mathcal M}\in \mathbb{C}^{M\times 1}$.
It is worth mentioning that ${\bf a}(\theta)$ in \eqref{EQU-4-1} is represented as a Kronecker product of the UPW-based far-field array response vectors of an $N$-element sparse ULA with inter-element spacing $\Gamma d$, i.e., ${\bf p}(\theta)$, and an $M$-element collocated ULA with inter-element spacing $d$, i.e., ${\bf b}(\theta)$. Fig. 2 illustrates the geometric relationship of such three array architectures, as well as their respective array response vectors.  \par
\textit{2) Sub-array based USW model for distinct AoAs/AoDs}: When $\frac{2S^2}{\lambda}\le r<\frac{2D^2}{\lambda}$, the user/scatterer is located within the far-field region of each array module, but within the near-field region of the whole array. As such, the array response vector ${\bf a}(r,\theta)$ in \eqref{EQU-3-1} can be simplified by exploiting the UPW model within each module $n$, as well as the USW model across different modules.
To this end, by letting $m=0$ in \eqref{EQU-1}, the distance from $\mathbf q$ to the reference element of module $n$ is
\begin{equation}\label{EQU-40} \vspace{-3pt}
\begin{split}
r_{n}\triangleq r_{n,0}=\sqrt{r^2-2r y_{n}\sin{\theta}+y_n^2}, \forall n\in\mathcal N,
\end{split}
\end{equation}
with $y_n=y_{n,0}=n\Gamma d$.
Additionally, let $\theta_n\in [-\frac{\pi}{2},\frac{\pi}{2}]$ represent the angle of user/scatterer $\mathbf q$ viewed from module $n$. Following from Fig. 1, we have
\begin{equation}\label{EQU-41} \vspace{-3pt}
\begin{split}
\sin\theta_n=\frac{r\sin\theta-y_n}{r_n}, \forall n\in\mathcal N.
\end{split}
\end{equation}
The array response vector ${\bf a}(r,\theta)$ in \eqref{EQU-3-1} thus simplifies as
\begin{equation}\label{EQU-5} \vspace{-3pt}
\begin{split}
&{\bf a}(r,\theta)=\left[e^{-j\frac{2\pi}{\lambda}r_n}{\bf b}(\theta_n)\right]_{n\in\mathcal N},\\
\end{split}
\end{equation}
where ${\bf b}(\theta_n)=\left[e^{j\frac{2\pi}{\lambda}md\sin\theta_{n}}\right]_{m\in\mathcal M}\in \mathbb{C}^{M\times 1}$.
In contrast to the UPW model in \eqref{EQU-4-1}, the near-field effect of ${\bf a}(r,\theta)$ in \eqref{EQU-5} is manifested in two aspects. Firstly, the phase variations across different modules are non-linear, since the exact distance $r_n$ is exploited to model the phase of the reference elements within each module, instead of its first-order Taylor approximation. Secondly, the AoAs/AoDs vary across different modules, since $\theta_n$ relates to the module index $n$.\par
\textit{3) Sub-array based USW model for common AoA/AoD}: When $\max\{5D, \frac{4SD}{\lambda}\}\le r< \frac{2D^2}{\lambda}$, we say that the user/scatterer is located within the extended far-field region of each
module, but within the near-field region of the whole array. The corresponding proof is provided in Appendix A in \cite{Lii2023}. We specify this extended far-field region by using the new distance criterion $\max\{5D, \frac{4SD}{\lambda}\}$, which is readily proved to be larger than the Rayleigh distance of each module $\frac{2S^2}{\lambda}$, since $D>\frac{S}{2}$. This leads to the angles $\theta_n$ with respect to different modules being approximately equal, i.e., $\theta_n \approx \theta, \forall n\in\mathcal N$. Hence, the array response vector ${\bf a}(r,\theta)$ in \eqref{EQU-5} further simplifies as
\begin{equation}\label{EQU-6} \vspace{-3pt}
\begin{split}
&{\bf a}(r,\theta)={\bf e}(r,\theta)\otimes {\bf b}(\theta),\\
\end{split}
\end{equation}
where
${\bf e}(r, \theta)=\left[e^{-j\frac{2\pi}{\lambda}r_{n}}\right]_{n\in\mathcal N}\in \mathbb{C}^{N\times 1}$.
It is different from the UPW model in \eqref{EQU-4-1} that ${\bf e}(r,\theta)$ in \eqref{EQU-6} is the USW-based near-field array response vector of an $N$-element sparse ULA with inter-element distance $\Gamma d$. Similar to \eqref{EQU-4-1}, the array geometric relationship of \eqref{EQU-6} is also described in Fig. 2.\par

\begin{figure}[htbp]
\centering
\includegraphics[scale=0.62]{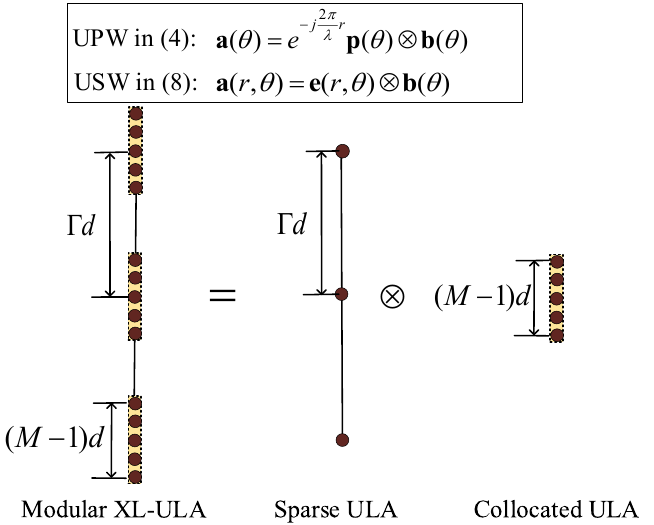}
\caption{The equivalent geometry of $NM$-element  modular XL-ULA under the UPW model and USW model with common AoA/AoD, which can be interpreted as the Kronecker product of an $N$-element sparse ULA and an $M$-element collocated ULA.}\label{movement}
  \vspace{-0.3cm}
\end{figure}

\section{Near-Field Beam Focusing Patterns and Grating Lobes for Modular XL-ULA} \label{model}\vspace{-1pt}
On basis of the near-field array models presented in Section II, we analyze the corresponding beam focusing patterns for modular XL-ULA in this section. Beam pattern characterization is crucial for multi-user communications, since it critically affects the level of IUI. For conventional UPW-based far-field array models, beam pattern provides the intensity variation of a beam designed for a certain direction as a function of the actual observation direction. However, for near-field models, the term \textit{beam focusing pattern} is used, since instead of direction only, it is the actual locations that completely specify the signal strength variations over array elements.
Specifically, for a beamforming vector $\mathbf v\in \mathbb{C}^{(N M)\times 1}$ of the modular XL-ULA designed for the desired location $(r',\theta')$, the beam focusing pattern at the near-field observation location $(r,\theta)$ is defined as
\begin{equation}\label{EQU-7} \vspace{-3pt}
\begin{split}
G(r,\theta;\mathbf v)&\triangleq \frac{1}{MN}{\left|\mathbf v^H{\bf a}(r,\theta)\right|},\\
\end{split}
\end{equation}
where ${\bf a}(r,\theta)$ denotes the near-field USW array response vector for the observation location $(r,\theta)$, which is modelled by (3).
Depending on the practical design and/or the acquired channel knowledge at the transmitter, both UPW-based far-field beamforming and USW-based near-field beamforming may be used for $\mathbf v$.  Thus, it is necessary to study the beam focusing patterns at the near-field observation location $(r,\theta)$ by using both far-field and near-field beamforming designs for $\mathbf v$. \par
Under far-field beamforming, the beamforming vector $\mathbf v$ only needs to match the desired direction $\theta'$ based on the UPW model in (4), i.e., ${\mathbf v}_{\rm FF}={\bf a}(\theta')$. By substituting ${\mathbf v}_{\rm FF}$ into (9), we have
\begin{equation}\label{EQU-71} \vspace{-3pt}
\begin{split}
&G_{\rm NF, FF}(r,\theta;\theta')\triangleq \frac{1}{MN}{\left|{\bf a}^H(\theta'){\bf a}(r,\theta)\right|}.\\
\end{split}
\end{equation}
Note that in the above, the subscript $``\rm {NF}"$ refers to near field for indicating the channel from the observation location $(r,\theta)$ based on the general near-field USW model, whereas the subscript $``\rm {FF}"$ refers to far field, thus indicating that the far-field UPW beamforming is used for $\mathbf v$. Similar notations are used in the sequel.
On the other hand, under near-field beamforming, $\mathbf v$ is designed to match the array response vector corresponding to the desired location $(r',\theta')$ based on the near-field USW model in (3), i.e.,  ${\mathbf v}_{\rm NF}={\bf a}(r',\theta')$. In this case, (9) is expressed as
\begin{equation}\label{EQU-72} \vspace{-3pt}
\begin{split}
G_{\rm NF, NF}(r,\theta;r',\theta')&\triangleq \frac{1}{MN}{\left|{\bf a}^H(r', \theta'){\bf a}(r,\theta)\right|}.\\
\end{split}
\end{equation}\par
In the following, we analyze the beam focusing patterns in (10) and (11) under far-field and near-field beamforming, respectively. To gain the essential insights, different scenarios of the observation point $(r,\theta)$ are considered by using the simplified array models for ${\bf a}(r,\theta)$ presented in Section II. \par
\subsection{Far-Field Observation with Far-Field Beamforming} \label{model}\vspace{-1pt}
When the observation location $(r,\theta)$ is  within the far-field region of the array, the near-field array response vector ${\bf a}(r,\theta)$ in (10) degenerates to (4). In this case, we have the following results.
\begin{theo}\label{theo1}
When $r\ge\frac{2D^2}{\lambda}$, $G_{\rm NF,FF}(r,\theta;\theta')$ in \eqref{EQU-71} degenerates to the beam pattern along the far-field observation direction with far-field beamforming, given by
\begin{equation}\label{EQU-100} \vspace{-3pt}
\begin{split}
G_{\rm FF,FF}(\theta;\theta')&=G_{\rm FF,FF}(\Delta_\theta)\\
&=\frac{1}{MN}\!\left|\sum_{n\in\mathcal N}\!\sum_{m\in\mathcal M}  e^{j\frac{2\pi}{\lambda}n\Gamma d\Delta_\theta}\!e^{j\frac{2\pi}{\lambda}md\Delta_\theta}\!\right|\\
&=\underbrace{\left|\frac{\sin \left(\pi N \Gamma \bar{d}\Delta_\theta\right)}{N\sin \left(\pi\Gamma \bar{d}\Delta_\theta\right)}\right|}_{\left|H_{N,\Gamma\bar{d}}(\Delta_\theta)\right|}\underbrace{\left|\frac{\sin \left(\pi M \bar{d}\Delta_\theta\right)}{M\sin \left(\pi \bar{d} \Delta_\theta\right)}\right|}_{\left|H_{M,\bar{d}}(\Delta_\theta)\right|},\\
\end{split}
\end{equation}
where $\Delta_\theta=\sin\theta-\sin\theta'$ denotes the difference of the spatial frequencies between the observation direction $\theta$ and the desired beamforming direction $\theta'$, ${\bar d}=\frac{d}{\lambda}$ is the inter-element spacing normalized by wavelength, and $H_{\tilde{M},\tilde{d}}(\Delta_\theta)\triangleq \frac{\sin \left(\pi \tilde{M}\tilde{d}\Delta_\theta\right)}{\tilde{M}\sin \left(\pi\tilde{d}\Delta_\theta\right)}$ is the Dirichlet kernel function, with the number of elements $\tilde{M}$ and normalized inter-element spacing $\tilde{d}$.
\begin{IEEEproof}
By substituting \eqref{EQU-4-1} into ${\bf a}(r,\theta)$ in \eqref{EQU-71}, Theorem 1 can be readily proved.
\end{IEEEproof}
\end{theo}\par

Theorem 1 shows that with far-field observation and far-field beamforming design, the beam pattern $G_{\rm FF,FF}(\Delta_\theta)$ only depends on the spatial frequency difference $\Delta_\theta$.  In addition, for the considered modular ULA, the beam pattern in \eqref{EQU-100} is expressed as a product of the far-field beam pattern of the sparse ULA with $N$ elements separated by the inter-element spacing $\Gamma d$, i.e.,  $|H_{N,\Gamma\bar{d}}(\Delta_\theta)|$, and that for the collocated ULA with $M$ elements separated by inter-element spacing $d$, i.e., $|H_{M,\bar{d}}(\Delta_\theta)|$.
Note that the result in \eqref{EQU-100} is consistent with the existing UPW beam pattern for sparse aperture radar array \cite{Zhuang2008}.

\begin{corollary}\label{lemma2}
By substituting  $\Gamma=M$,  $G_{\rm FF,FF}(\theta;\theta')$ in \eqref{EQU-100} reduces to that of the conventional collocated ULA \cite{2022}, i.e.,
\begin{equation}\label{EQU-111} \vspace{-3pt}
\begin{split}
G_{\rm FF,FF}(\theta;\theta')=G_{\rm FF,FF}(\Delta_\theta)=\underbrace{\left|\frac{\sin \left(\pi MN \bar{d}\Delta_\theta\right)}{MN\sin \left(\pi \bar{d}\Delta_\theta\right)}\right|}_{\left|H_{MN,\bar{d}}(\Delta_\theta)\right|}.\\
\end{split}
\end{equation}
\end{corollary}\par
Corollary 1 shows that the developed far-field beam pattern for modular ULA includes the conventional collocated ULA as a special case.
To obtain more insights, by analyzing the function $|H_{\tilde M, \tilde d}(\Delta_\theta)|$, some important properties of far-field beam patterns for modular and collocated ULAs in \eqref{EQU-100} and  \eqref{EQU-111} are given below.\par

{\it 1) Spatial Angular Resolution:} By letting  $\pi \tilde{M}\tilde{d} \Delta_\theta=\pm \pi$ or $ \Delta_\theta=\pm \frac{1}{\tilde{M}\tilde{d}}$, we have
$|H_{\tilde M, \tilde d}(\Delta_\theta)|=0$.
Therefore, the null-to-null beam width of the main lobe of $|H_{\tilde M, \tilde d}(\Delta_\theta)|$ can be obtained as $\frac{2}{\tilde{M}\tilde{d}}$ \cite{222}. Thus, the null-to-null beam widths of the main lobes for the two terms $|H_{N, \Gamma \bar d}(\Delta_\theta)|$ and $|H_{M, \bar d}(\Delta_\theta)|$ in  \eqref{EQU-100} are  $\frac{2}{N\Gamma \bar d}$ and $\frac{2}{M \bar d}$, respectively.
Since $ N\Gamma \ge M$, the overall null-to-null beam width of the main
lobe for modular XL-ULA in (12) is determined by ${\left|H_{N,\Gamma\bar{d}}(\Delta_\theta)\right|}$ and thus  given by $\frac{2}{N\Gamma\bar{d}}$. Typically, angular resolution can be defined as  half of the main
lobe beam width \cite{333}. Thus, the spatial angular resolution of the considered modular ULA is
\begin{equation}\label{EQU-11} \vspace{-3pt}
\begin{split}
{\Delta}^{\mathrm{mod}}_{\theta, \rm res}=\frac{1}{N\Gamma\bar{d}},
\end{split}
\end{equation}
which depends on the aperture of the entire array.
As a comparison, the angular resolution of the collocated ULA in \eqref{EQU-111} can be similarly obtained as ${\Delta}^{\mathrm{col}}_{\theta, \rm res}=\frac{1}{N M\bar{d}}$. Since $\Gamma \ge M$, this shows the superiority of modular array to its collocated array counterpart in improving the angular resolution.
As an illustration, Fig. 3 shows the far-field beam patterns for modular and collocated ULAs with equal number of elements, demonstrating the higher angular resolution of modular ULA than its collocated ULA counterpart.\par
{\it 2) Grating Lobes:}  For the function $|H_{\tilde M, \tilde d}(\Delta_\theta)|$, when the inter-element spacing is larger than half of the signal wavelength, i.e., $\tilde{d}>\frac{1}{2}$, grating lobes appear. Specifically, by letting $\pi \tilde{d}\Delta_\theta= i\pi$, $i=\pm1,\pm2,\cdots$ \cite{222}, the angular separation of each grating lobe is obtained as $\Delta_\theta= i \frac{1}{\tilde{d}}, i=\pm1,\pm2,\cdots$.
As such, since $\Delta_\theta\in[-2,2]$, the total number of grating lobes is $\lfloor 4\tilde{d} -1\rfloor$ for $\tilde{d}>\frac{1}{2}$. Therefore, for a fixed number of array elements, while increasing the element separation $\tilde{d}$ enhances the spatial resolution, it also leads to more grating lobes.
For the considered modular ULA, since $\Gamma \bar{d}>\frac{1}{2}$, the term $\left|H_{N,\Gamma\bar{d}}(\Delta_\theta)\right|$ in \eqref{EQU-100} results in grating lobes that are separated by
\begin{equation}\label{EQU-110} \vspace{-3pt}
\begin{split}
\Delta_\theta=\frac{1}{\Gamma\bar{d}},
\end{split}
\end{equation}
which may cause angular ambiguity.
Fig. 3 shows the grating lobes of far-field UPW beam patterns for modular ULA. It is observed that compared to its collocated ULA counterpart with inter-element separation $\bar{d}=\frac{1}{2}$,
 modular array generates  undesired grating lobes with a period of $\frac{1}{\Gamma \bar d}$ appearing in the angular domain. Fortunately, it is also observed that for modular array, undesired grating lobes can be suppressed to certain extent by the term $\left|H_{M,\bar{d}}(\Delta_\theta)\right|$ in \eqref{EQU-100}, which serves as the envelope for $G_{\rm FF,FF}(\Delta_\theta)$.\par

\begin{figure}[htbp]
\centering
\includegraphics[scale=0.44]{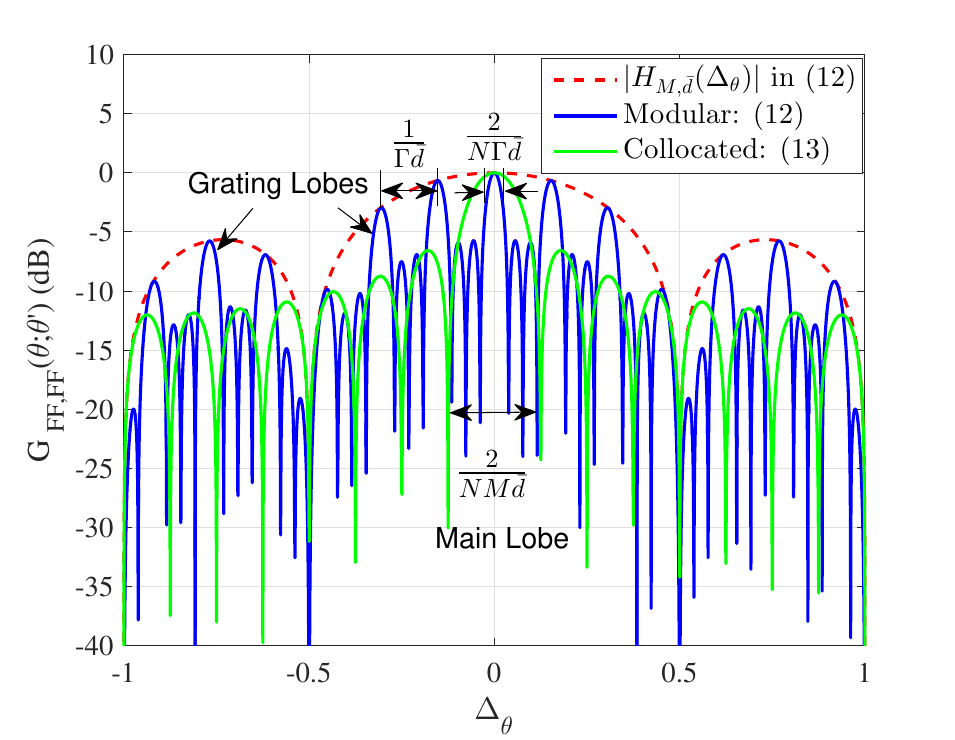}
\caption{The beam patterns along the far-field observation direction with far-field beamforming for modular and collocated ULAs, both having $16$ elements. For modular ULA, the number of modules is $N=4$, the number of antennas in each module is $M=4$, and the modular size is $\Gamma=13$.}\label{movement}
  \vspace{-0.3cm}
\end{figure}
\subsection{Near-Field Observation with Far-Field Beamforming} \label{model}\vspace{-1pt}
When the observation location $(r,\theta)$ is within the Rayleigh distance of the modular XL-ULA, we use the two near-field array models developed in Section II to simplify the beam focusing pattern in (10).

\begin{theo}\label{theo2}
When $\frac{2S^2}{\lambda}\le r<\frac{2D^2}{\lambda}$, the beam focusing pattern at the near-field observation location by using far-field beamforming $G_{\rm NF, FF}(r,\theta;\theta')$ in \eqref{EQU-71} can be expressed as
\begin{equation}\label{EQU-162}  \vspace{-0.2cm}
\begin{split}
&G_{\rm NF, FF}(r,\theta;\theta')\\
&=\frac{1}{MN}\left|\sum_{n\in\mathcal N}\!\sum_{m\in\mathcal M} \! e^{-j\frac{2\pi}{\lambda}(r_n+n\Gamma d\sin\theta')}\!e^{j\frac{2\pi}{\lambda}md(\sin\theta_n-\sin\theta')}\!\right|\\
&\!=\frac{1}{N}\!\Bigg |\!\sum_{n\in\mathcal N} \! e^{-j\frac{2\pi}{\lambda}(r_n+n\Gamma d\sin\theta')}\!\underbrace{\frac{\sin \left(\pi M \bar{d}(\sin\theta_n\!-\sin\theta')\right)}{M\sin \left(\pi \bar{d} (\sin\theta_n\!-\sin\theta')\right)}}_{H_{M,\bar{d}}(\sin\theta_n\!-\sin\theta')}\!\Bigg |.\\
\end{split}
\end{equation}
\begin{IEEEproof}
By substituting ${\bf a}(r,\theta)$ in (10) with (7), the proof is established.
\end{IEEEproof}
\end{theo}\par
Theorem 2 shows that different from the far-field beam pattern in \eqref{EQU-100} that only depends on the spatial frequency difference $\Delta_\theta$, the beam focusing pattern at the near-field observation location under far-field beamforming in \eqref{EQU-162} depends on the actual observation location $(r,\theta)$ and the desired beamforming direction $\theta'$.\par
\begin{theo}\label{theo2}
When $\max\{5D, \frac{4SD}{\lambda}\}\le r< \frac{2D^2}{\lambda}$, $G_{\rm NF, FF}(r,\theta;\theta')$ in \eqref{EQU-162} can be expressed as
\begin{equation}\label{EQU-172}  \vspace{-0.2cm}
\begin{split}
&G_{\rm NF, FF}(r,\theta;\theta')=\frac{1}{N}\Bigg |\sum_{n\in\mathcal N}  e^{-j\frac{2\pi}{\lambda}(r_n+n\Gamma d\sin\theta')}\Bigg|\Big|{H_{M,\bar{d}}(\Delta_{\theta})}\Big|.\\
\end{split}
\end{equation}
\begin{IEEEproof}
Under the condition of Theorem 3, by letting $\sin \theta_n \approx \sin\theta$, $\forall n \in \mathcal N$, the result in (17) can be obtained from \eqref{EQU-162}.
\end{IEEEproof}
\end{theo}\par

Fig. 4 plots the beam focusing pattern of modular ULA by using far-field beamforming based on the original expression in (10), as well as its simplified expressions in (16) and (17). It is observed that the three curves match quite well with each other, demonstrating the effectiveness of the developed sub-array based USW models for
distinct versus common AoAs/AoDs with respect to different sub-arrays/modules in Section II.
It is also observed from Fig. 4 that if the far-field beamforming design is mistakenly used in the near-field observation environment, the beam focusing patterns can cause the expansion of beam widths, termed \textit{energy spread effect} or \textit{power leakage effect} \cite{Gao2022, You2023, Dai2022}. For the newly considered modular array, the power leakage effect will be more severe than its collocated array counterpart due to the existence of grating lobes and its larger total aperture, as evident by comparing the green curve in Fig. 4 with the other three curves.

\begin{figure}[htbp]
\centering
\includegraphics[scale=0.42]{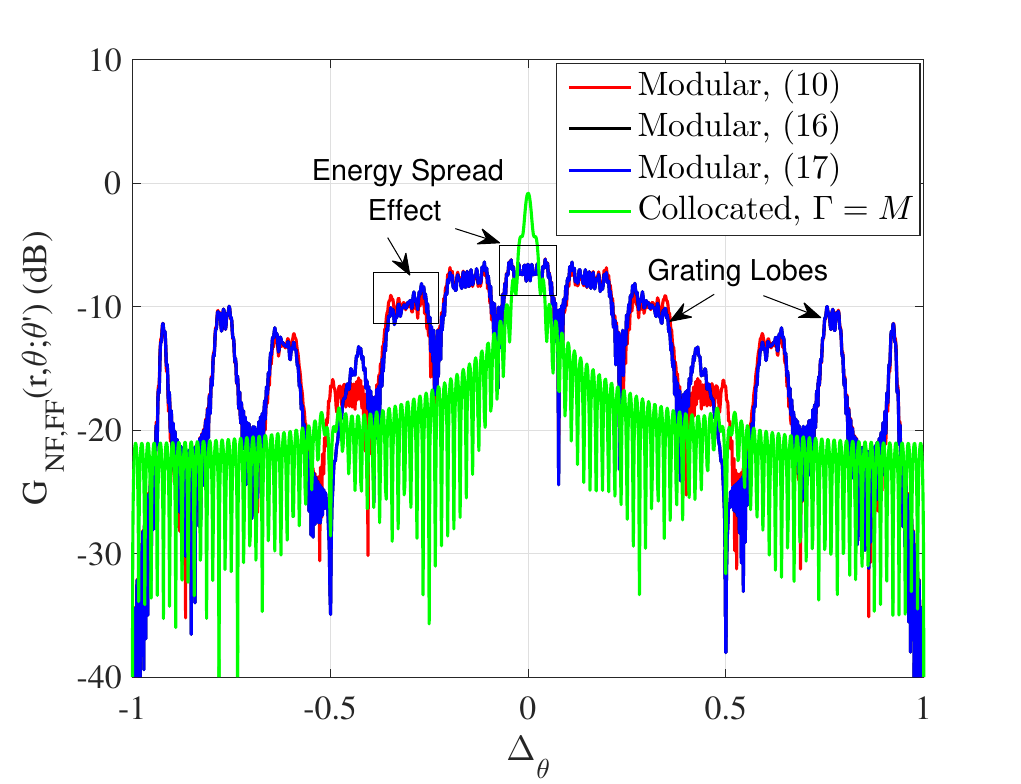}
\caption{The beam focusing patterns at the near-field observation location with far-field beamforming for modular and collocated XL-ULAs, both having $128$ elements. For modular XL-ULA, the number of antennas in each module is $M=32$, the number of modules is $N=4$, and the modular size is $\Gamma=13$. The desired beam focusing location is $(r',\theta')=($200$ $\!\!\!\!\!\quad\rm m$, 0)$, and the observation distance is $r=200$ $\rm m$.}\label{movement}
  \vspace{-0.3cm}
\end{figure}\par

\subsection{Near-Field Observation with Near-Field Beamforming} \label{model}\vspace{-1pt}

In this subsection, we consider the beam focusing pattern at the near-field observation location by using near-field beamforming in (11).

\begin{theo}\label{theo2}
When $\frac{2S^2}{\lambda}\le r<\frac{2D^2}{\lambda}$, the beam focusing pattern at the near-field observation location under near-field beamforming $G_{\rm NF,NF}(r,\theta;r', \theta')$ in \eqref{EQU-72} can be expressed as
\begin{equation}\label{EQU-12}  \vspace{-0.2cm}
\begin{split}
G_{\rm NF,NF}(r,\theta;r',\theta')&=\frac{1}{MN}\left|\sum_{n\in\mathcal N}\!\sum_{m\in\mathcal M}  e^{-j\frac{2\pi}{\lambda}\Delta_{r,n}}\!e^{j\frac{2\pi}{\lambda}md\Delta_{\theta,n}}\!\right|\\
&=\frac{1}{N}\Bigg |\sum_{n\in\mathcal N}  e^{-j\frac{2\pi}{\lambda}\Delta_{r,n}}\underbrace{\frac{\sin \left(\pi M \bar{d}\Delta_{\theta,n}\right)}{M\sin \left(\pi \bar{d} \Delta_{\theta,n}\right)}}_{H_{M,\bar{d}}(\Delta_{\theta,n})}\Bigg |,\\
\end{split}
\end{equation}
where $\Delta_{r,n}=r_n-r_n'=\sqrt{r^2-2r y_{n}\sin{\theta}+y_n^2}-\sqrt{r'^2-2r' y_{n}\sin{\theta'}+y_n^2}$, and $\Delta_{\theta,n}=\sin\theta_n-\sin\theta_n'=\frac{r\sin\theta-y_n}{r_n}-\frac{r'\sin\theta'-y_n}{r_n'}, \forall n\in\mathcal N$.
\begin{IEEEproof}
By substituting \eqref{EQU-5} into \eqref{EQU-72}, the proof is established.
\end{IEEEproof}
\end{theo}\par
Theorem 4 shows that the near-field beam focusing pattern for modular XL-ULA  in \eqref{EQU-12} depends on the locations $(r_n,\theta_n)$ and $(r_n',\theta_n')$ viewed from different array modules. It is expressed in the form of a weighted sum of $N$ beam patterns, each corresponding to one module of $M$-element ULAs and having different spatial frequency differences $\Delta_{\theta,n}$.  The complex-weighted coefficient $e^{-j\frac{2\pi}{\lambda}\Delta_{r,n}}$ is dependent on the accurate phase difference with respect to the center element of module $n$. Since $\Delta_{r,n}$ and $\Delta_{\theta,n}$ are correlated with the module index $n$ in a sophisticated way, the closed-form expression for \eqref{EQU-12} is difficult to obtain. However, when the observation distance $r$ is moderately large, the expression in \eqref{EQU-12} can be further simplified in the following.



\begin{theo}\label{theo2}
When $\max\{5D, \frac{4SD}{\lambda}\}\le r< \frac{2D^2}{\lambda}$, $G_{\rm NF,NF}(r,\theta;r', \theta')$ in \eqref{EQU-72} can be expressed as
\vspace{-0.1cm}
\begin{equation}\label{EQU-132} \vspace{-3pt}
\begin{split}
G_{\rm NF,NF}(r,\theta;r',\theta')=\frac{1}{N}\left|\sum_{n\in\mathcal N} e^{-j\frac{2\pi}{\lambda}\Delta_{r,n}}\right|\left|H_{M,\bar{d}}(\Delta_\theta)\right|.\\
\end{split}
\end{equation}
\begin{IEEEproof}
Under the condition of Theorem 5, by considering $\sin \theta_n \approx \sin\theta$ and $\sin\theta_n'\approx \sin\theta'$, $\forall n \in \mathcal N$, the result in (19) can be obtained from \eqref{EQU-12}.
\end{IEEEproof}
\end{theo}\par
Theorem 5 shows that based on the sub-array based USW model for common AoA/AoD, the near-field beam focusing pattern can be separated into two parts.
The first one denotes the USW beam focusing pattern for the sparse ULA having $N$ elements, while the second part represents the UPW beam pattern for the
collocated ULA having $M$ elements. Furthermore,
as the number of modules $N$ becomes large, a closed-form expression for \eqref{EQU-132} can be achieved in the following.
\begin{corollary}\label{theo2}
When $N$ becomes large, $G_{\rm NF,NF}(r,\theta;r',\theta')$ in \eqref{EQU-132} can be obtained as \eqref{eq6} in closed-form, as presented at the top of the next page,
\begin{figure*}
\begin{equation}\vspace{-3pt}
\label{eq6}
G_{\rm NF,NF}(r,\theta;r',\theta')\!=\!\left\{
\begin{aligned}
&\!\left|\frac{F\left(\sqrt{| \nu_{r,\theta}|}\frac{N}{2}+\frac{\mu_{\theta}}{2\sqrt{| \nu_{r,\theta}|}}\right)+F\left(\sqrt{|\nu_{r,\theta}|}\frac{N}{2}-\frac{\mu_{\theta}}{2\sqrt{|\nu_{r,\theta}|}}\right)}{\sqrt{|\nu_{r,\theta}|}N}\right|{\left|H_{M,\bar{d}}(\Delta_\theta)\right|} & , \quad \! \frac{\cos^2\theta}{r}\neq \frac{\cos^2\theta'}{r'}, \\
&{\left|H_{N,\Gamma\bar{d}}(\Delta_\theta)\right|}{\left|H_{M,\bar{d}}(\Delta_\theta)\right|}
& , \quad \!  \frac{\cos^2\theta}{r}= \frac{\cos^2\theta'}{r'},\\
\end{aligned}
\right.
\end{equation}
{\noindent} \rule[0pt]{17.5cm}{0.05em}
  \vspace{-0.3cm}
\end{figure*}
where  $\nu_{r,\theta}=-{\pi\bar d}\Gamma^2 d\delta_{r,\theta}$, $\mu_\theta={2\pi \bar d}\Gamma\Delta_{\theta}$, and $\delta_{r,\theta}=\frac{\cos^2\theta}{r}-\frac{\cos^2\theta'}{r'}$. Besides, $F(x)=C(x)+jS(x)$, with $C(x)= \int_{0}^{x} \cos t^{2} \mathrm{~d} t$ and $S(x)=\int_{0}^{x} \sin t^{2} \mathrm{~d} t$ being Fresnel integrals \cite{Gradshteyn2007}.
\end{corollary}\par
\begin{pproof}Please refer to Appendix B in \cite{Lii2023}. $\hfill \blacksquare$\end{pproof}  \par
Corollary 2 shows that the expression in \eqref{eq6} relates to the angle-distance difference $\delta_{r,\theta}$, and spatial frequency difference  $\Delta_\theta$. By adopting the distance ring $\xi$, the curve $r=\xi\cos^2\theta$ is defined in polar coordinate \cite{Dai2022}. Obviously, when $(r,\theta)$ and $(r',\theta')$ are located at the same distance ring $\xi$, i.e., $\frac{1}{\xi}=\frac{\cos^2\theta}{r}=\frac{\cos^2\theta'}{r'}$, the USW-based and UPW-based results respectively given in \eqref{eq6} and \eqref{EQU-100} are identical. For a common scenario when locations are at two different distance rings, the near-field beam focusing pattern is dependent on the function $F(x)$ with respect to Fresnel integrals. \par




\begin{figure}[htbp]
\centering
\includegraphics[scale=0.51]{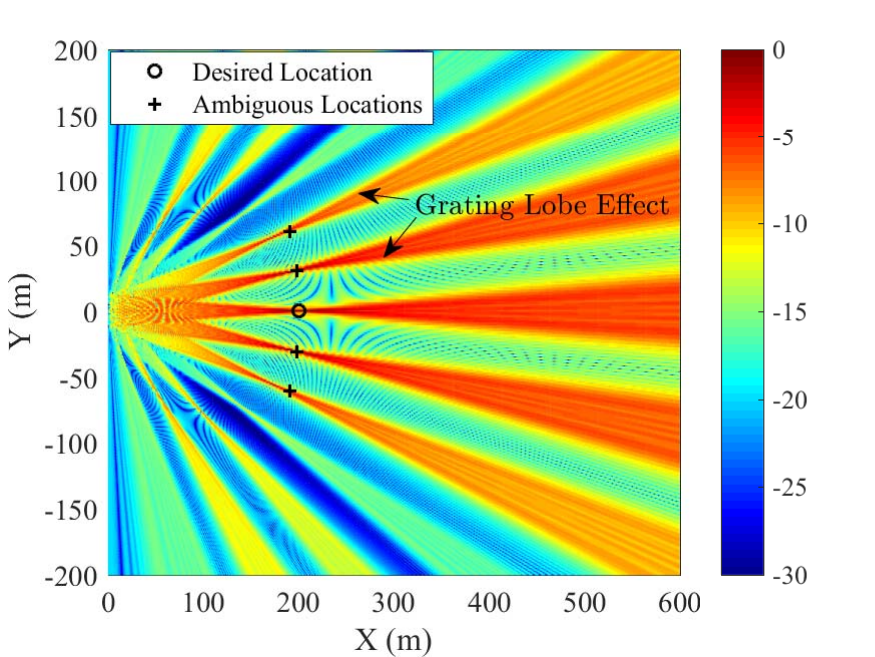}
\caption{The near-field grating lobe effect of modular XL-ULA. The parameters of modular XL-ULA are same as Fig. 4 and the desired beam focusing location is $(r',\theta')=($200$ $\!\!\!\!\!\quad\rm m$, 0)$, labelled as black circle in the figure.}\label{movement}
  \vspace{-0.3cm}
\end{figure}

Fig. 5 plots the near-field beam focusing pattern of the same modular array considered in Fig. 4, which clearly shows the near-field grating lobe effect that may lead to undesired ambiguous locations, besides the desired location.  To gain further insights, some important properties of the USW-based near-field beam focusing patterns for modular XL-ULA  in \eqref{EQU-132} and \eqref{eq6} are provided as follows.\par
{\it 1) Spatial Resolution:}
Compared to the conventional UPW-based far-field model in (12) that only supports angular resolution, the USW-based near-field model associated with modular XL-ULA can provide the spatial resolution from both angular and distance dimensions. However, since the non-linear phase varies across different modules, the closed-form beam width of the main lobe in either angular or distance domain is difficult to be directly derived. In this case, we use half of the approximated null-to-null main lobe beam width in the angular domain as the effective angular resolution, and half of the half-power ($-3$  $\rm dB$) main lobe
beam width in the distance domain as the effective distance resolution.\par
Specifically, when the two locations have the same distance but different directions, i.e., $r=r'$ and $\theta\neq \theta'$, the effective angular resolution of modular XL-ULA under near-field beamforming is approximated as that of the UPW-based far-field model in (12), i.e.,
${\Delta}^{\mathrm{mod}}_{\theta,\rm res}=\frac{1}{N \Gamma \bar d}$.

On the other hand, when the two locations are along the same direction while having different distances, i.e., $\theta=\theta'$ and $r\neq r'$, we have $\mu_\theta=0$ and $\nu_{r, \theta}=-{\pi\bar d}\Gamma^2 d\cos^2\theta'\left( \frac{1}{r}-\frac{1}{r'}  \right)$. In this case, the expression in (20) can be simplified as
\begin{equation}\label{EQU-1300} \vspace{-3pt}
\begin{split}
G_{\rm NF,NF}(r,\theta';r',\theta')=\left|\frac{F\left(\frac{N\Gamma|\cos\theta'|}{2}\sqrt{{\pi\bar d} d \left|\frac{1}{r}-\frac{1}{r'} \right|}\right)}{\frac{N\Gamma|\cos\theta'|}{2}\sqrt{{\pi\bar d} d \left|\frac{1}{r}-\frac{1}{r'} \right|}}\right|.
\end{split}
\end{equation}\par
By defining $z\triangleq \frac{1}{r}-\frac{1}{r'}$, $G_{\rm NF,NF}(r,\theta';r',\theta')$ in \eqref{EQU-1300} can be rewritten as
\begin{equation}\label{EQU-130} \vspace{-3pt}
\begin{split}
E_{\rm NF,NF}\!\left(z;\theta'\!\right)=\left|\!\frac{F\left(\frac{N\Gamma|\cos\theta'|}{2}\sqrt{{\pi\bar d} d \left|z \right|}\right)}{\frac{N\Gamma|\cos\theta'|}{2}\sqrt{{\pi\bar d} d \left|z \right|}}\!\right|,
\end{split}
\end{equation}
which is an even function with respect to $z$.
As such, we first provide the definition of the effective distance resolution of modular XL-ULA under near-field beamforming with respect to $z$ as
\begin{equation}\label{EQU-7231}
\begin{split}
{\Delta}^{\rm mod}_{1/r,\rm res}(\theta')&\triangleq{\rm argmin}_{z>0} \Big\{E_{\rm NF,NF}\!\left(z;\theta'\!\right)\le \frac{1}{2}\Big\}.\\
\end{split}
\end{equation}\par
Note that the effective distance resolution in terms of $z$ depends on the desired beamforming direction $\theta'$ in general.
\begin{theo}
The effective distance resolution of modular XL-ULA with respect to $z$ in \eqref{EQU-7231} is
\begin{equation}\label{EQU-22216} \vspace{-3pt}
{\Delta}^{\rm mod}_{1/r,\rm res}(\theta')=\frac{1}{r_{\rm hp}(\theta')},
\end{equation}
where $r_{\rm hp}(\theta')=0.10 \cos^2\theta'\frac{2 D^2}{\lambda}$ is defined as \textit{the half-power effective distance}, and $D\approx \Gamma Nd$ is the total array aperture.
\end{theo}
\begin{pproof}Please refer to Appendix A. $\hfill \blacksquare$\end{pproof}  \par
Theorem 6 shows that the effective distance resolution depends on the half-power effective distance $ r_{\rm hp}(\theta')$, which further depends on the total aperture of the modular XL-ULA, $D$ and the direction, $\theta'$.
It is also shown that the half-power effective distance $r_{\rm hp}(\theta')$ is smaller than the classical Rayleigh
distance $r_{\rm rayl}=\frac{2D^2}{\lambda}$ \cite{122} and the effective Rayleigh distance $r_{\rm eff}(\theta')=0.367 \cos^2\theta'\frac{2 D^2}{\lambda}$ \cite{2223}.

Moreover, in terms of $\Delta_r \triangleq r-r'$, $G_{\rm NF,NF}(r,\theta';r',\theta')$ in \eqref{EQU-1300} can be also rewritten as
\begin{equation}\label{EQU-13001} \vspace{-3pt}
\begin{split}
G_{\rm NF,NF}\!(r'\!+\!\Delta_r,\theta';r',\theta'\!)\!=\!\left|\!\frac{F\!\left(\frac{N\Gamma|\cos\theta'|}{2}\sqrt{{\pi\bar d} d \left|\frac{\Delta_r}{\Delta_r r'+r'^2}\right|}\!\right)}{\frac{N\Gamma|\cos\theta'|}{2}\sqrt{{\pi\bar d} d \left|\frac{\Delta_r}{\Delta_r r'+r'^2}\right|}}\!\right|.
\end{split}
\end{equation}\par
Note that different from (22), $G_{\rm NF,NF}\!(r'\!+\!\Delta_r,\theta';r',\theta'\!)$ in \eqref{EQU-13001} with respect to $\Delta_r$ is not symmetric about $\Delta_r =0$. Therefore, the effective distance resolution with respect to  ${\Delta}_r$ should be separately defined for two cases as follows.
For $\Delta_r>0$, the effective distance
resolution with respect to  ${\Delta}_r$ is defined as
\begin{equation}\label{EQU-22} \vspace{-3pt}
\begin{split}
\!{\Delta}^{\rm mod}_{\!r^+,\rm res}(r'\!,\theta')\!\triangleq\!
{\rm argmin}_{\!\Delta_r>0}\! \Big\{G_{\rm NF,NF}\Big({r'}\!+\!\Delta_r,\!\theta';\!r',\!\theta'\Big)\!\le \frac{1}{2}\Big\}, \\
\end{split}
\end{equation}
while for $\Delta_r<0$, we have
\begin{equation}\label{EQU-22} \vspace{-3pt}
\begin{split}
\!{\Delta}^{\rm mod}_{r^-,\rm res}\!(r'\!,\theta')\!\triangleq\!
-\!{\rm argmax}_{\Delta_r\!<0}\! \Big\{G_{\rm NF,NF}\!\Big({r'}\!+\!\Delta_r,\!\theta';\!r',\!\theta'\!\Big)\!\le \! \frac{1}{2}\Big\}. \\
\end{split}
\end{equation}\par
Note that the effective distance resolutions with respect to ${\Delta}_r$ in (26) and (27) depend on the
desired beam focusing location $(r', \theta')$ in general.
\begin{theo}
The effective distance
resolution with respect to $\Delta_r$ in (26) is
\begin{equation}\label{EQU-222171} \vspace{-3pt}
{\Delta}^{\rm mod}_{r^+,\rm res}(r',\theta')=
\left\{
\begin{aligned}
&\frac{r'^2}{r_{\rm hp}(\theta')-r'}, \quad\quad r'<r_{\rm hp}(\theta'),\\
&\infty ,\quad\quad\quad\quad\quad\quad\textit{otherwise},\\
\end{aligned}
\right.
\end{equation}
while for $\Delta_r<0$, the effective distance
resolution in (27) is
\begin{equation}\label{EQU-222172} \vspace{-3pt}
{\Delta}^{\rm mod}_{r^-,\rm res}(r',\theta')= \frac{r'^2}{r_{\rm hp}(\theta')+r'}.
\end{equation}
\end{theo}
\begin{pproof}Please refer to Appendix B. $\hfill \blacksquare$\end{pproof}  \par

It can be inferred from Theorem 7 that since ${\Delta}^{\rm mod}_{r^+,\rm res}(r',\theta')>{\Delta}^{\rm mod}_{r^-,\rm res}(r',\theta')$, modular XL-ULA can provide better distance resolution when the user is located closer to the array, as expected. Moreover, the overall half-power
beam width of the main lobe for modular XL-ULA under near-field beamforming in the distance domain is obtained as ${\rm BW}^{\rm mod}_r={\Delta}^{\rm mod}_{r^+,\rm res}(r',\theta')+{\Delta}^{\rm mod}_{r^-,\rm res}(r',\theta')$.

\begin{figure}[htbp]
\centering
 \includegraphics[scale=0.4]{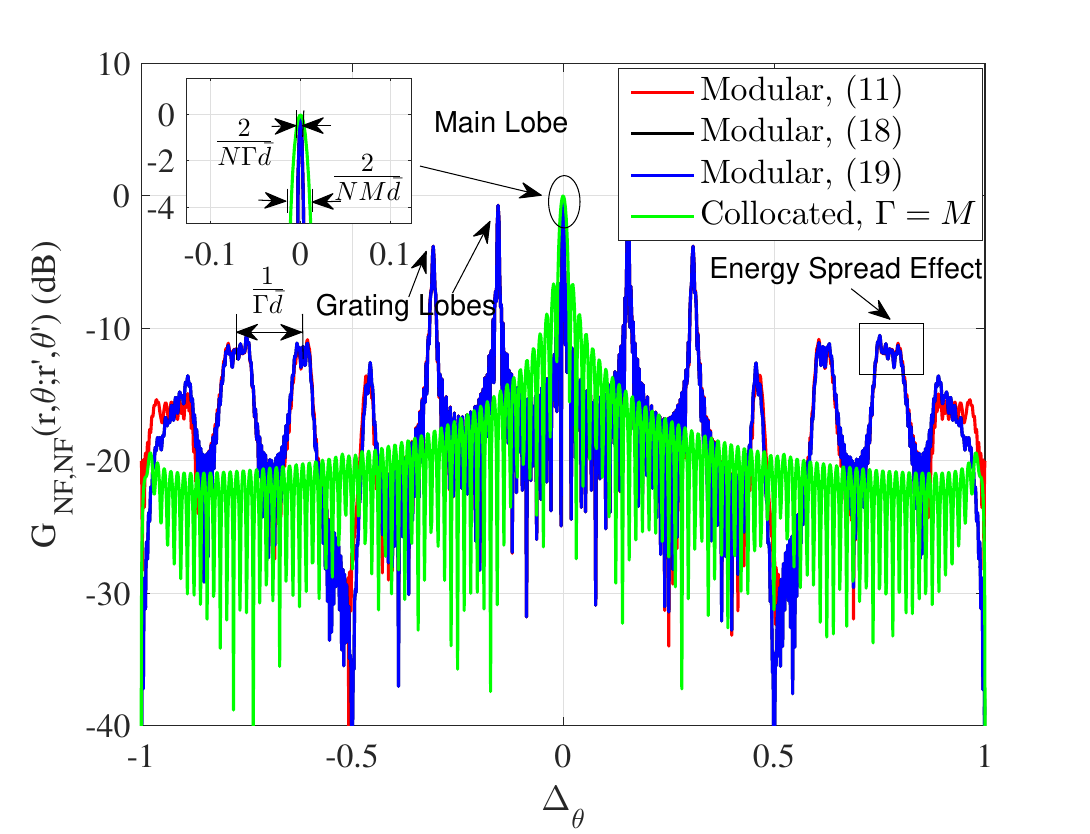}
\caption{The beam focusing patterns at the near-field observation location with near-field beamforming versus spatial frequency differences $\Delta_\theta$, where the desired beam focusing location is $(r',\theta')=($200$ $\!\!\!\!\!\quad\rm m$, 0)$, and the observation distance is $r=200$ $\rm m$. The parameters of modular and collocated XL-ULAs are same as Fig. 4.}\label{movement}
\end{figure}

\begin{figure}[htbp]
\centering
 \includegraphics[scale=0.49]{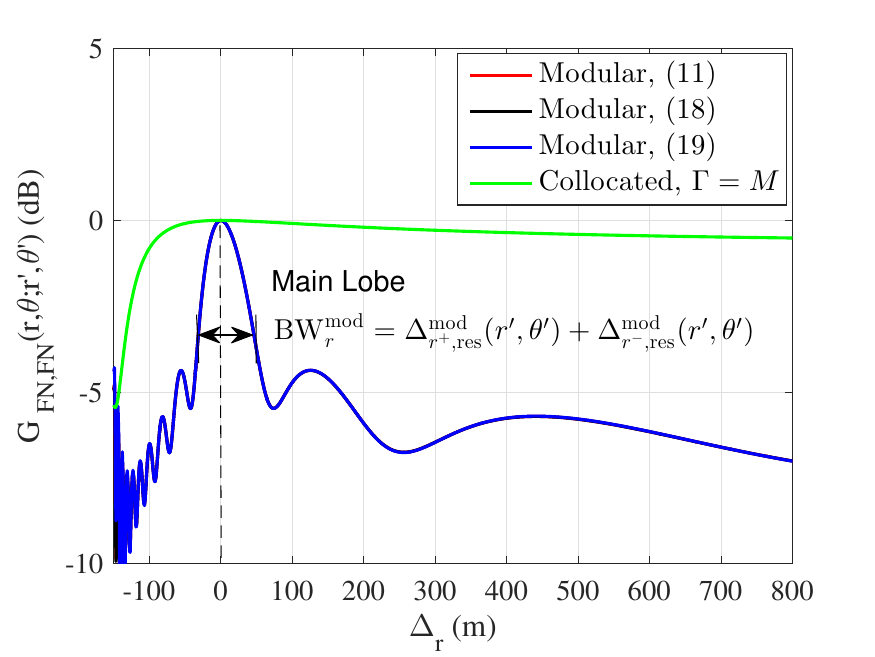}
\caption{The beam focusing patterns at the near-field observation location with near-field beamforming versus distance separations $\Delta_{r}$, where the desired beam focusing location is $(r',\theta')=($200$ $\!\!\!\!\!\quad\rm m$, 0)$, and the observation direction is $\theta=0$. The  parameters of modular and collocated XL-ULAs are same as Fig. 4.}\label{movement}
\vspace{-0.3cm}
\end{figure}

To gain more insights, the asymptotic performance limits of the effective distance resolution are revealed as follows. Specifically, as $D$ becomes infinitely large, $r_{\rm hp}(\theta') \to \infty$ except for the singular direction $\theta'=\frac{\pi}{2}$. In this case, both \eqref{EQU-222171} and \eqref{EQU-222172} become
\begin{equation}\label{EQU-22211}
\lim_{D\to\infty}{\Delta}^{\rm mod}_{r^+,\rm res}(r',\theta')=\lim_{D\to\infty}{\Delta}^{\rm mod}_{r^-,\rm res}(r',\theta')=0.
\end{equation}
This indicates that with the USW-based near-field model, continuously increasing the array aperture will lead to extremely refined distance resolution.
On the other hand, when $r'$ increases to infinity, the results in both \eqref{EQU-222171} and \eqref{EQU-222172}
reduce to
\begin{equation}\label{EQU-22212}
\lim_{r'\to\infty}{\Delta}^{\rm mod}_{r^+,\rm res}(r',\theta')=\lim_{r'\to\infty}{\Delta}^{\rm mod}_{r^-,\rm res}(r',\theta')=\infty,
\end{equation}
which indicates that there is no distance resolution when the user/scatterer is far away from the array, as expected.

Fig. 6 and Fig. 7 illustrate the near-field beam focusing patterns for modular array and collocated array architectures  versus  spatial frequency differences and distance variations, respectively. It is observed that modular XL-ULA can provide better spatial resolution from both angular and distance dimensions as compared to its collocated counterpart, but may generate undesired grating lobes.\par
{\it 2) Grating Lobes:}
For the considered near-field beam focusing patterns of modular XL-ULA, grating lobes periodically occur with their approximated center directions $\Delta_\theta=i\frac{1}{\Gamma \bar d}$, $i=\pm1,\pm2,....$. By comparing with the far-field beamforming design in Fig. 4, the proper near-field beamforming design shown in Fig. 6 not only better suppresses grating lobes, but also achieves higher spatial resolution.\par


\section{User Grouping for Multi-User Modular XL-MIMO Communications}
The analysis of beam focusing patterns in Section III reveals that undesired grating lobes may cause severe IUI for multi-user modular XL-MIMO systems. Fortunately, this issue can be addressed by proper user grouping for multi-user transmission scheduling, i.e., users located within
the grating lobes of each other are not assigned
to the same time-frequency communication RB for their communications, as elaborated below.

\subsection{Near-Field Channel Model for Multi-User Modular XL-MIMO Communications}

Section II presents a generic near-field array response of modular XL-ULA for a user/scatterer located at $(r,\theta)$. Based on such array responses, in this section, we consider an uplink multi-user modular XL-MIMO communication system in multi-path setting. Let $K$ denote the number of single-antenna users, and $L_k+1$ denote the number of channel paths for user $k$, including one possible line-of-sight (LoS) component and $L_k$ non-line-of-sight (NLoS) components. Let  $(r_{k,0},\theta_{k,0})$ denote the location of user $k$, with $k$ belonging to the user set ${\mathcal K}=\{1,..,K\}$, and $(r_{k,l},\theta_{k,l})$, $1\leq l \leq L_k$,  denote the location of its $l$-th scatterer.
Based on the array response vectors presented in Section II, the near-field channel vector for user $k$ under the multi-path setting can be modelled as
\begin{equation}\label{EQU-20} \vspace{-3pt}
{\bf h}^{\rm NF}_{k}=\sum_{l=0}^{L_k}\alpha_{k,l} {\bf a}(r_{k,l},\theta_{k,l}),
\end{equation}
where $\alpha_{k,l}$ and ${\bf a}(r_{k,l},\theta_{k,l})\in \mathbb{C}^{(NM)\times 1}$ respectively represent the complex-valued coefficient and near-field array response vector of the $l$-th path of user $k$, with $l=0$ corresponding to the LoS path. For the special case of far-field scenario, the multi-path channel in \eqref{EQU-20} degenerates to
\begin{equation}\label{EQU-21} \vspace{-3pt}
{\bf h}^{\rm FF}_k=\sum_{l=0}^{L_k}\alpha_{k,l} {\bf a}(\theta_{k,l}),
\end{equation}
where ${\bf a}(\theta_{k,l})\in \mathbb{C}^{(NM)\times 1}$ represents the far-field array response vector of the $l$-th path of user $k$.
We assume that the $K$ users communicate with the modular XL-ULA by sharing $Q$ orthogonal time-frequency RBs, where $Q<K$.
Across the $Q$ RBs, the channels of all users are assumed to remain unchanged. A binary variable $x_{q,k}\in \{0,1\}$ is introduced, $q\in {\mathcal Q}=\{1,..,Q\}$, where
\begin{equation}\label{EQU-22z} \vspace{-3pt}
x_{q,k}=\left\{
\begin{aligned}
1 & , \quad\textit{user k is allocated to resource q,} \\
0 & , \quad\textit{otherwise}.\\
\end{aligned}
\right.
\end{equation}\par
Accordingly, for each RB $q$, the users belonging to the set ${\mathbb S}_q=\{k|x_{q,k}=1, \forall k\in \mathcal K\}$ are grouped to transmit over it.

The received signal by the modular XL-ULA at RB $q$ can be expressed as
\begin{equation}\label{EQU-24} \vspace{-3pt}
\begin{split}
&{{\bf y}_q}=
\sum_{i\in {\mathbb S}_q}{\bf h}^{\rm NF}_{i} {\sqrt P_i}s_i+{\bf n}_q, \forall q\in \mathcal Q,\\
\end{split}
\end{equation}
where $P_i$ is the transmit power by user $i$, and $s_i$ with $\mathbb{E}[|s_k|^2]=1$ is the informating-bearing signal of user $i$. Furthermore, ${\bf n}_q$ denotes the additive white Gaussian noise (AWGN), following the CSCG distribution with power $\sigma^2$, i.e., ${\bf n}_q\sim{\cal CN}(0,{\sigma}^2{\bf I}_{N M })$.\par

For user $k \in {\mathbb S}_q$, let ${\bf v}_k\in\mathbb{C}^{(N M )\times 1}$ with $||{\bf v}_k||=1$ denote the receive beamforming vector. Then, the resulting signal for user $k$ is
\begin{equation}\label{EQU-24} \vspace{-3pt}
\begin{split}
&{y_{q,k}}=
{\bf v}_k^H{\bf h}^{\rm NF}_{k} \sqrt{P_k}s_k+\sum_{i\in {\mathbb S}_q,i\neq k}{\bf v}_k^H{\bf h}^{\rm NF}_{i} {\sqrt P_i}s_i+{\bf v}_k^H {\bf n}_q.\\
\end{split}
\end{equation}

As a result, the SINR for user $k$ is
\begin{equation}\label{EQU-25} \vspace{-3pt}
\gamma_{k}({\bf v}_k,{\mathbb S}_q)=\frac{\bar{P}_k|{\bf v}_k^H{\bf h}^{\rm NF}_{k}|^2}{\sum_{i\in {\mathbb S}_q,i\neq k}\bar{P}_i|{\bf v}_k^H{\bf h}^{\rm NF}_{i}|^2+1}, \forall k\in {\mathbb S}_q,
\end{equation}
where $\bar{P}_i=\frac{P_i}{{\sigma}^2}$ denotes the transmit SNR for user $i$.

The achievable sum rate for the $K$ users over the $Q$ RBs in bits/second/Hz (bps/Hz) is
\begin{equation}\label{EQU-26} \vspace{-3pt}
R=\sum_{q\in \mathcal Q}\sum_{k\in {\mathbb S}_q}{\rm \log}_2(1+\gamma_{k}({\bf v}_k,{\mathbb S}_q)).
\end{equation}\par
By jointly optimizing the beamforming vectors ${\bf v}_k$ and user grouping index matrix $[x_{q,k}]_{Q\times K}$,
the problem for maximizing the achievable sum rate can be formulated as
\begin{equation}\label{EQU-611} \vspace{-3pt}
\begin{split}
&\max_{{\bf v}_k, [x_{q,k}]_{Q\times K}}  R\\
&\text{s.t.} ~ C1: ||{\bf v}_k||=1, \forall k \in \mathcal K,\\
&~\quad C2:  \sum_{q\in \mathcal Q} x_{q,k}=1, \forall k \in \mathcal K,\\
&~\quad C3: x_{q,k}\in\{0,1\}, \forall k \in \mathcal K, q \in \mathcal Q,\\
\end{split}
\end{equation}
where  the constraint $C2$ represents that each user is allocated to one RB only. Problem \eqref{EQU-611} is a mixed-integer non-convex optimization problem, thus it is
challenging to obtain its globally optimal solution efficiently. To find an efficient suboptimal solution, we alternatively optimize the beamforming vectors ${\bf v}_k$ by adopting the classic linear beamforming methods, and the user grouping index matrix $[x_{q,k}]_{Q\times K}$ by leveraging the greedy method.

\subsection{Beamforming with Fixed User Grouping} \label{model}\vspace{-1pt}
We first consider the three classical beamforming methods for problem (39), by assuming that the user grouping is given for multi-user transmission scheduling. For communicating with users located in the near-field region, the conventional model-based channel state information (CSI) acquisition methods, such as the compressed sensing or codebook-based beam sweeping, may render the BS acquire the mismatched far-field CSI as in (29) if the far-field model is presumed. Therefore, in this subsection, both the far-field and near-field based linear beamforming are considered. With far-field beamforming, the beamforming vectors $\mathbf{v}_k$ are designed based on the far-field CSI $\hat{\mathbf{h}}_{k}={\mathbf{h}^{\rm FF}_k}$, whereas for the near-field beamforming, we have $\hat{\mathbf{h}}_{k}={\mathbf{h}^{\rm NF}_k}$. Therefore, unless otherwise stated, the following schemes are applicable to both near-field and far-field beamforming, by using the corresponding CSI $\hat{\mathbf{h}}_{k}$.

\subsubsection{MRC beamforming} \label{model}\vspace{-1pt}
\par
With the low complexity MRC beamforming for user $k$, i.e., $\mathbf{v}_{k}^{\mathrm{MRC}}=\frac{\hat {\mathbf{h}}_k}{\left\|\hat{\mathbf{h}}_k\right\|}$, the SINR in
\eqref{EQU-25} can be written as
\begin{equation}\label{EQU-51} \vspace{-3pt}
\begin{split}
\gamma_{k}(\mathbf{v}_{k}^{\mathrm{MRC}},{\mathbb S}_q)&=\frac{ \bar{P}_k\frac{\left|\hat{\mathbf{h}}_k^H{\mathbf{h}^{\rm NF}_k}\right|^2}{\|\hat{\mathbf{h}}_k\|^2}}{\sum_{i\in{\mathbb S}_q, i \neq k} \bar{P}_i \frac{\left|\hat{\mathbf{h}}_k^H {\mathbf{h}^{\rm NF}_i}\right|^2}{\left\|\hat{\mathbf{h}}_k\right\|^2}+1}.\\
\end{split}
\end{equation}
For the special case of $\hat{ \mathbf{h}}_{k}={\mathbf{h}^{\rm NF}_k}$, the SINR in \eqref{EQU-51} reduces to
\begin{equation}\label{EQU-512} \vspace{-3pt}
\begin{split}
\gamma_{k}(\mathbf{v}_{k}^{\mathrm{MRC}},{\mathbb S}_q)=\bar{P}_k\left\|{\mathbf{h}^{\rm NF}_k}\right\|^2\left(1-\beta^{\mathrm{MRC}}_{k}\right),\\
\end{split}
\end{equation}
where $\beta^{\mathrm{MRC}}_{k}=\frac{\sum_{i\in \mathbb{S}_q, i \neq k} \bar{P}_i \rho_{k i}^2\left\|{\mathbf{h}^{\rm NF}_i}\right\|^2}{\sum_{i\in \mathbb{S}_q, i \neq k} \bar{P}_i \rho_{k i}^2\left\|{\mathbf{h}^{\rm NF}_i}\right\|^2+1}$ is the SNR loss factor, which increases with $\rho_{k i}$, and $\rho_{ki}=\frac{\left|(\mathbf{h}^{\rm NF}_k)^{H} {\mathbf{h}^{\rm NF}_i}\right|}{\left\|{\mathbf{h}^{\rm NF}_k}\right\|\left\|{\mathbf{h}^{\rm NF}_i}\right\|}$ is the correlation between the near-field channel vectors for user $k$ and user $i$.
For the LoS-only scenario, $\rho_{ki}$ is simplified as
\begin{equation}\label{EQU-511} \vspace{-3pt}
\begin{split}
\rho_{ki}=\frac{1}{MN}\left|{\bf a}^H(r_{k},\theta_{k}){\bf a}(r_{i},\theta_{i})\right|,
\end{split}
\end{equation}
which can be interpreted as the near-field beam
focusing pattern when
considering $(r_k,\theta_k)$ as the desired beam focusing location, and $(r_i,\theta_i)$ as the
observation location. Therefore, the beam focusing pattern presented in Section III can be directly applied for our SINR analysis.
\subsubsection{ZF beamforming} \label{model}\vspace{-1pt}
With ZF beamforming, we have $\mathbf{v}^{\mathrm{ZF}}_{k}=\frac{(\mathbf{I}_{N M}-\hat{\mathbf{A}}_k) {\hat{\mathbf{h}}_k}}{\left\|(\mathbf{I}_{N M }-\hat{\mathbf{A}}_k){\hat{\mathbf{h}}_k}\right\|}$, where
$\hat{\mathbf{A}}_k=\hat{\mathbf{B}}_k\left((\hat{\mathbf{B}}_k)^H \hat{\mathbf{B}}_k\right)^{-1} (\hat{\mathbf{B}}_k)^H$ and
$\hat{\mathbf{B}}_k=\left[{\hat{\mathbf{h}}_i}\right] \in\mathbb{C}^{(N M)\times (N_q-1)}$,  $i\neq k$ \cite{Brown2012}. The resulting SINR is
\begin{equation}\label{EQU-513w} \vspace{-3pt}
\begin{split}
\gamma_{k}(\mathbf{v}_{k}^{\mathrm{ZF}},{\mathbb S}_q)&=\frac{ \bar{P}_k\frac{\left|\hat{\mathbf{h}}_k(\mathbf{I}_{N M}-
\hat{\mathbf{A}}_k)^H{\mathbf{h}^{\rm NF}_k}\right|^2}{\|(\mathbf{I}_{N M}-\hat{\mathbf{A}}_k)\hat{\mathbf{h}}_k\|^2}}{\sum_{i\in{\mathbb S}_q, i \neq k} \bar{P}_i \frac{\left|\hat{\mathbf{h}}_k^H(\mathbf{I}_{N M }-\hat{\mathbf{A}}_k)^H {\mathbf{h}^{\rm NF}_i}\right|^2}{\left\|(\mathbf{I}_{N M}-\hat{\mathbf{A}}_k) \hat{\mathbf{h}}_k\right\|^2}+1}.\\
\end{split}
\end{equation}
For the special case of $\hat{ \mathbf{h}}_{k}={\mathbf{h}^{\rm NF}_k}$,
the SINR in \eqref{EQU-513w} degenerates to
\begin{equation}\label{EQU-514} \vspace{-3pt}
\begin{split}
\gamma_{k}(\mathbf{v}_{k}^{\mathrm{ZF}},{\mathbb S}_q)= \bar{P}_k\left\|\mathbf{h}^{\rm NF}_{k}\right\|^{2}\left(1-\beta^{\mathrm{ZF}}_{k}\right),
\end{split}
\end{equation}
where $\beta^{\mathrm{ZF}}_{k}=\frac{(\mathbf{h}^{\rm NF}_k)^H \mathbf{A}_k {\mathbf{h}^{\rm NF}_k}}{\left\|{\mathbf{h}^{\rm NF}_k}\right\|^2}$,
${\mathbf{A}}_k={\mathbf{B}}_k\left({\mathbf{B}}_k^H {\mathbf{B}}_k\right)^{-1} ({\mathbf{B}}_k)^H$ and
${\mathbf{B}}_k=\left[{\mathbf{h}^{\rm NF}_i}\right] \in\mathbb{C}^{(N M)\times (N_q-1)}$,  $i\neq k$. If the RB $q$ only accommodates two users $k$ and $i$, we have $\beta^{\mathrm{ZF}}_{k}=\rho_{k i}^2$, which also increases with $\rho_{k i}$.\par
\subsubsection{MMSE beamforming} \label{model}\vspace{-1pt}
With MMSE beamforming, we have $\mathbf{v}^{\mathrm{MMSE}}_{k}=\frac{\hat{\mathbf{C}}_k^{-1} {\hat{\mathbf{h}}_k}}{\left\|\hat{\mathbf{C}}_k^{-1} {\hat{\mathbf{h}}_k}\right\|}$, where $\hat{\mathbf{C}}_k=\sum_{i\in {\mathbb S}_q, i \neq k}\bar{P}_i {\hat{\mathbf{h}}_i} {\hat{\mathbf{h}}_i}^H+\mathbf{I}_{ N M }$.  The SINR in \eqref{EQU-25} is
\begin{equation}\label{EQU-515} \vspace{-3pt}
\begin{split}
\gamma_{k}(\mathbf{v}_{k}^{\mathrm{MMSE}},{\mathbb S}_q)&=\frac{ \bar{P}_k\frac{\left|\hat{\mathbf{h}}_k^H(\hat{\mathbf{C}}_k^{-1})^H{\mathbf{h}^{\rm NF}_k}\right|^2}{\|\hat{\mathbf{C}}_k^{-1}\hat{\mathbf{h}}_k\|^2}}{\sum_{i\in{\mathbb S}_q, i \neq k} \bar{P}_i \frac{\left|\hat{\mathbf{h}}_k^H(\hat{\mathbf{C}}_k^{-1})^H {\mathbf{h}^{\rm NF}_i}\right|^2}{\left\|\hat{\mathbf{C}}_k^{-1} \hat{\mathbf{h}}_k\right\|^2}+1}.\\
\end{split}
\end{equation}
For the special case of $\hat{ \mathbf{h}}_{k}={\mathbf{h}^{\rm NF}_k}$, \eqref{EQU-515} reduces to
\begin{equation}\label{EQU-516} \vspace{-3pt}
\begin{split}
\gamma_k(\mathbf{v}_{k}^{\mathrm{MMSE}},{\mathbb S}_q)= \bar{P}_k\left\|\mathbf{h}_{k}^{\rm NF}\right\|^{2}\left(1-\beta^{\mathrm{MMSE}}_{k}\right),
\end{split}
\end{equation}
where $\beta^{\mathrm{MMSE}}_{k}=\frac{(\mathbf{h}^{\rm NF}_k)^H(\mathbf{I}_{N M }-\mathbf{C}_k^{-1}) {\mathbf{h}^{\rm NF}_k}}{\left\|{\mathbf{h}^{\rm NF}_k}\right\|^2}$, and $\mathbf{C}_k=\sum_{i\in {\mathbb S}_q, i \neq k}\bar{P}_i {\mathbf{h}^{\rm NF}_i} (\mathbf{h}^{\rm NF}_i)^H+\mathbf{I}_{ N M }$.
If there are only two users $k$ and $i$ in the RB $q$, we have
$\beta^{\mathrm{MMSE}}_{k}=\frac{\bar{P}_i\left\|{\mathbf{h}^{\rm NF}_i}\right\|^{2} \rho_{ki}^2}{1+\bar{P}_i\left\|{\mathbf{h}^{\rm NF}_i}\right\|^{2}}$ which again increases with $\rho_{k i}$.\par

It is found that the SINRs under three linear beamforming schemes highly depend on the beam focusing pattern, and can be degraded by the pronounced grating lobes. This shows the importance to address the grating lobe issue caused by the modular XL-array architecture.

\subsection{User Grouping with Fixed Beamforming} \label{model}\vspace{-1pt}

\begin{algorithm}[tbp]
\caption{Greedy algorithm for user grouping}\label{table1}
\centering
\begin{tabular}{ll}  
1: \textbf{Initialize}  ${\mathbb T}_{u}=\{1,...,K\}$,  and ${\mathbb S}_{q}=\emptyset$, $\forall q\in \mathcal Q$;\\
2: \textbf{While} ${\mathbb T}_{u}\neq \emptyset$ do:\\
3: \quad\textrm{Randomly pick up a user} $k$ from ${\mathbb T}_{u}$;\\
4: \quad \textbf{Let} $R_{\max}=0$; \\
5: \quad \textbf{for} $q\in \mathcal Q$; \\
6: \quad\quad \textrm{Calculate} $R_{q}=\sum_{i\in{\mathbb S}'_{q}}\log_2(1+\gamma_{{i}}(\mathbf{v}_{i}^*, {\mathbb S}'_{q}))$,\\
   \quad\quad \quad\! where $\!{\mathbb S}'_{q}\!=\!{\mathbb S}_{q}\cup \{k\}\!$ \textrm{and} $\mathbf{v}_{i}^*$ is the various\\
   \quad\quad \quad\! beamforming obtained in Section IV-B;\\
7:  \quad \quad \textrm{Calculate} $\!{\hat R}_{\rm sum}\!\!=\!\!R_{q}\!\!+\!\!\sum_{\!j\neq q,\!j\in \mathcal Q}\! \!\sum_{\!i\in{\mathbb S}_j}\!\log_2\!(1\!+\!\gamma_{i}\!(\!\mathbf{v}_{\!i}^*\!,\! {\mathbb S}_{j}\!)\!)$;\\
8: \quad \quad \textbf{If} ${\hat R}_{\rm sum}>R_{\max}$;\\
9: \quad \quad \quad\textrm{Let} $R_{\max}={\hat R}_{\rm sum}$, and $q^*=q$;\\
10:\quad\quad\!\textbf{End};\\
11: \quad\!\textbf{End};\\
12: \quad\!${\mathbb S}_{q^*}={\mathbb S}_{q^*}\cup\{k\}$, and  ${\mathbb T}_{u}={\mathbb T}_{u}/\{k\}$;\\
13: \textbf{End};\\
14: \textbf{Output} $R=R_{\max}$ and ${\mathbb S}_{q}$, $\forall q\in \mathcal Q$.\\
\end{tabular}
\end{algorithm}

For the given beamforming $\mathbf{v}_{k}^*$ presented in the previous subsection, for designing the user grouping index matrix $[x_{q,k}]_{Q\times K}$, the problem (39) reduces to
\begin{equation}\label{EQU-61} \vspace{-3pt}
\begin{split}
&\max_{[x_{q,k}]_{Q\times K}}\sum_{q\in \mathcal Q}\sum_{k\in {\mathbb S}_q}{\rm \log}_2(1+\gamma_{k}(\mathbf{v}_{k}^*, {\mathbb S}_{q}))\\
&\text{s.t.} ~ C2,C3.\\
\end{split}
\end{equation}\par
However, the problem \eqref{EQU-61} is a mixed-integer non-convex problem, for which it is difficult to obtain a globally optimal solution efficiently. Therefore, we propose an efficient greedy
algorithm to solve problem \eqref{EQU-61} sub-optimally, as summarized in Algorithm 1.
Specifically, in step 3, we first randomly pick up a user $k$ from the unallocated user index set ${\mathbb T}_{u}$. Then, if user $k$ is allocated into the RB $q$,
we calculate the achievable sum rate for the $q$-th RB as $R_q$, and the corresponding sum rate, $\hat{R}_{\rm sum}$. Note that $\mathbf{v}_{i}^*$ corresponds to the beamforming presented in Section IV-B. After comparing all possible allocations $q$, the best RB $q^*$ leading to the maximum sum rate is selected. The process continues until all the $K$ users are allocated over the $Q$ RBs. The overall complexity of the greedy based user grouping algorithm is of order $\mathbb {O}(QK)$.\par

\section{Simulation Results} \label{model}\vspace{-1pt}
As shown in Fig. 8, we consider a multi-user modular XL-MIMO system in a hot-spot area, where $K$ users are scattered in a disk region, with its center $(r_c,0)$ and radius $r_{\rm max}<r_c$. As such, the distance range of users is $r_k\in [r_c-r_{\rm max},r_c+r_{\rm max}]$, and the angular range of the users is $\theta_{k,0}\in [-\theta_{\rm max},\theta_{\rm max}]$, with $\theta_{\rm max}=\arcsin\left(\frac{r_{\rm max}}{r_c}\right)$.
For multi-path scenarios, the location $(r_{k,l},\theta_{k,l})$ of the $l$-th scatterer of user $k$ follows the distributions $r_{k,l} \sim {\cal U}(0, 200 \!\!\!\quad\! \!\rm m)$ and $\theta_{k,l} \sim {\cal U}(-\frac{\pi}{2}, \frac{\pi}{2})$.

Unless otherwise specified, the number of RBs is $Q=15$, the number of users is $K=30$, the number of NLoS paths of user $k$ is $L_k=20$, the circular radius is $r_{\rm max}=20$ $\rm m$, and the distance from the BS to the center of circle is $r_{c}=200$ $\rm m$.  The total number of elements for antenna array is $NM=128$. For modular array, the number of modules is $N=32$, the number of antennas within each module is $M=4$, and modular size parameter is $\Gamma=13$. The inter-element distance is $d=\frac{\lambda}{2}=0.0628$ $\rm m$, corresponding to the frequency of $2.38$ $\rm GHz$. Thus, with the transmit SNR for each user $\bar{P}_k=\bar{P}_t=90$ $\rm dB$, the reference-element receive SNR is $\bar{P}_r=\bar{P}_t\left(\frac{\lambda}{4\pi r_c}\right)^2\approx 2.50$ $\rm dB$. For the multi-path channel, the channel gain for the LoS path of user $k$ is $\alpha_{k,0}=\frac{\lambda}{4\pi r_{k,0}}$, while
that for the $l$-th NLoS path of user $k$ is $\alpha_{k,l}=\frac{\lambda \sqrt{\sigma_{k,l}}}{(4\pi)^{3/2}t_{k,l} r_{k,l}}e^{-j\frac{2\pi}{\lambda}t_{k,l}+jw_{k,l}}$ \cite{Dong2022}, where $t_{k,l}$ is the distance from scatterer $l$ to the user $k$, $w_{k,l}$ is the phase shift following from uniform distribution over $[-\pi, \pi)$ and $\sigma_{k,l}$ represents the radar cross
section (RCS) of scatterer $l$ belonging to uniform distribution over $[1,40]$ $\rm m^2$.\par

\begin{figure}[htbp]
\centering
\includegraphics[scale=0.58]{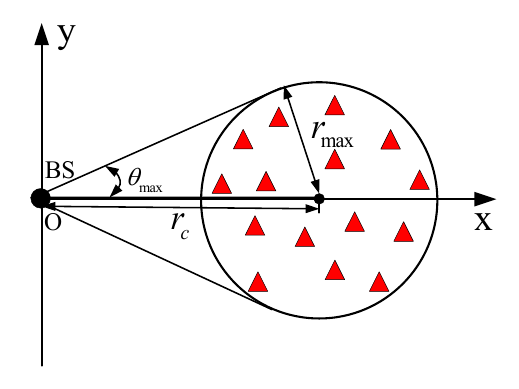}
\caption{Geometric illustration of user distribution.}\label{movement}
 \vspace{-0.3cm}
\end{figure}

\begin{figure}[htbp]
\centering
\includegraphics[scale=0.5]{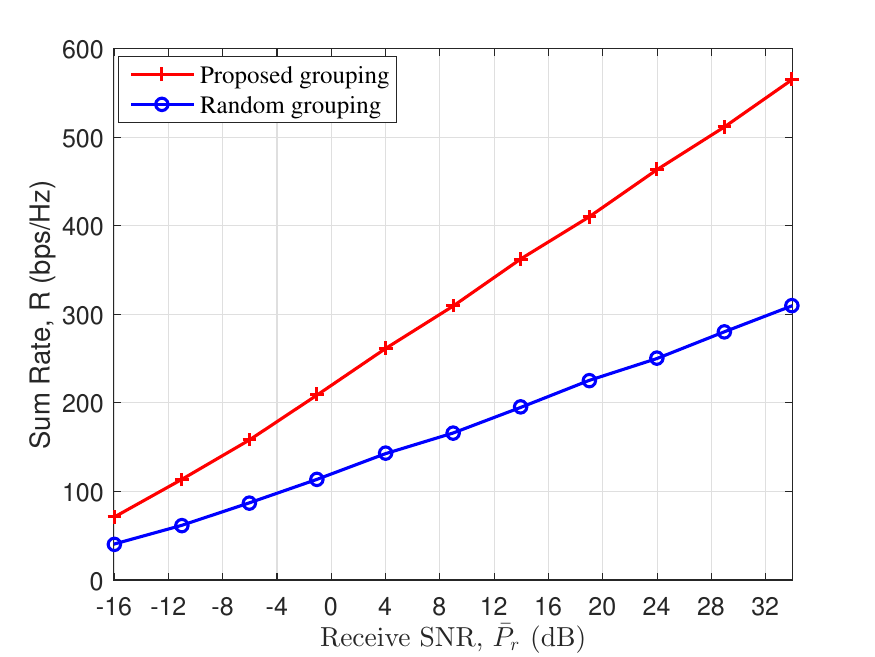}
\caption{Achievable sum rates for modular XL-MIMO communications with proposed and random user grouping.}\label{movement}
\vspace{-0.3cm}
\end{figure}

\par
Fig. 9 plots the achievable sum rate, $R$ under the multi-path scenario versus the receive SNR, ${\bar P}_r$, with near-field MMSE beamforming. It is observed that the proposed user grouping algorithm can achieve significantly larger sum rates than random user grouping.
This demonstrates the effectiveness of the greedy based user grouping method to mitigate the grating lobe induced IUI in  modular XL-MIMO communications. \par

\begin{figure}[t]
  \centering{
    \label{1} 
    \includegraphics[scale=0.5]{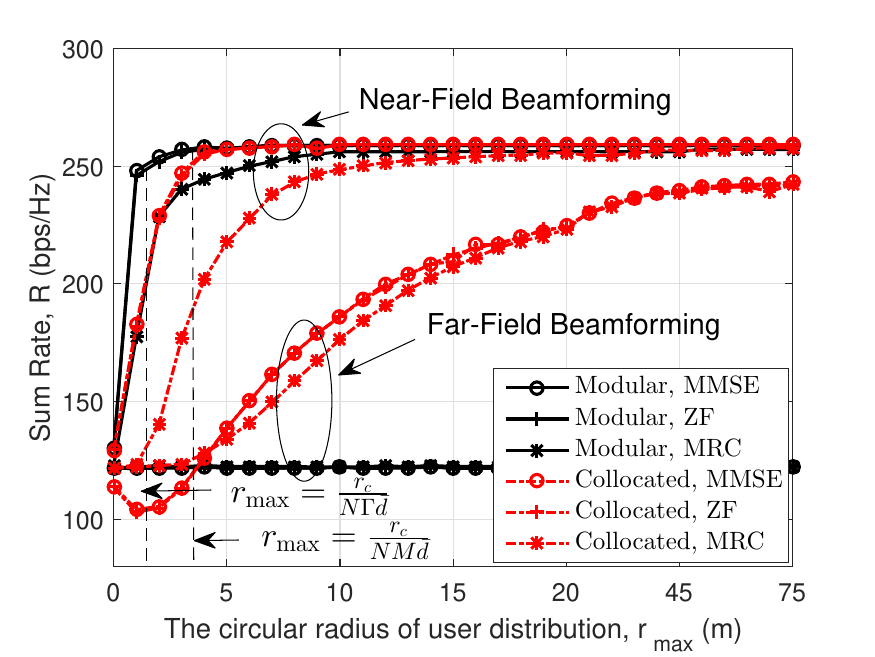}}
    \hspace{0in}
  \caption{Achievable sum rates for modular and collocated arrays versus the circular radius of user distribution, under far-field and near-field linear beamforming.}\label{pic1}
  \label{12}
  \vspace{-0.3cm}
\end{figure}

Fig. 10 plots the achievable sum rates for modular and collocated arrays versus the circular radius, $r_{\rm max}$ in user
distribution, according to the far-field and near-field linear beamforming designs, including MRC, ZF and MMSE.
It is observed that for the three near-field linear beamforming designs, when the circular radius of user
distribution is relatively small, e.g., $r_{\rm max}\in \left[\frac{r_c}{N \Gamma \bar{d}},\frac{r_c}{N M \bar{d}}\right]$, the achievable sum rates for modular XL-MIMO are much larger than the collocated counterpart by benefiting from its higher spatial resolution from both angular and distance dimensions. When the circular radius of user
distribution becomes large enough, the achievable sum rates for both array architectures tend to be comparable.
Moreover, by comparing the far-field and near-field beamforming, we see that the near-field beamforming achieves significantly larger rate than the far-field beamforming.
These observations verify the necessity of near-field beamforming for near-field scenarios, and the superiority of modular XL-MIMO over the collocated XL-MIMO for communicating with densely distributed users. \par

\begin{figure}[t]
  \centering
    \includegraphics[scale=0.5]{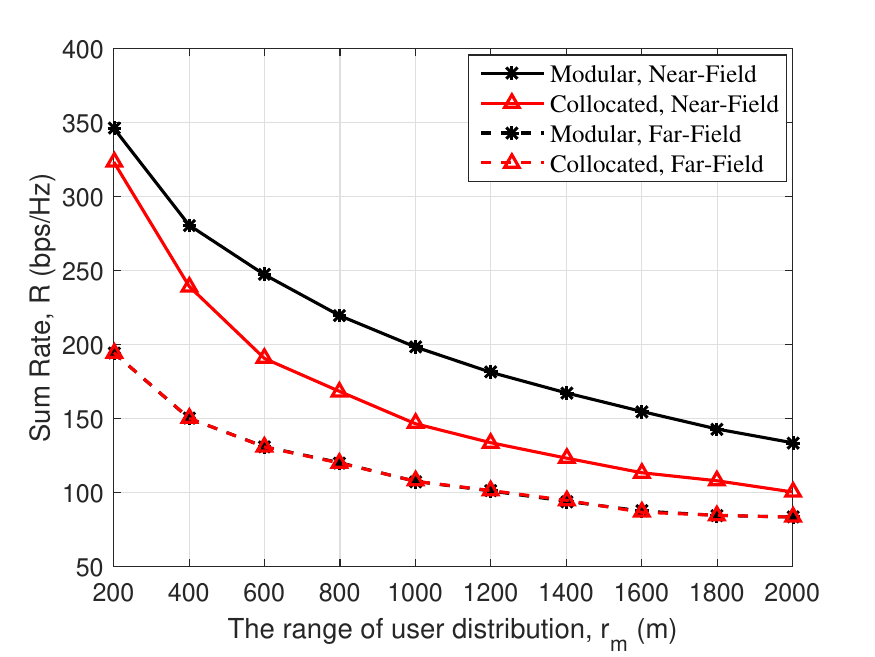}
    \hspace{0in}
  \caption{Achievable sum rates for different array architectures with MMSE beamforming versus the range of user distribution.}\label{pic1}
  \label{12}
  \vspace{-0.3cm}
\end{figure}

Lastly, we consider another scenario, where $K$ users are assumed to randomly distribute along the positive direction of $x$-axis over the distance range $[0,r_m]$. Fig. 11 plots the achievable sum rates for different array architectures with MMSE beamforming versus the user
distance range, $r_m$. It is observed that with the increase of distance range, the achievable sum rates for both array architectures decline, as expected. More importantly, we observe that the near-field beamforming for modular XL-MIMO can significantly enhance the sum rates much more than its collocated counterpart. However, the two rate curves of the far-field beamforming overlap with each other. This is expected since only the near-field beamforming can provide the effective spatial resolution over the distance dimension.\par

\section{Conclusions}\label{g} \vspace{-1pt}
In this paper, we studied multi-user modular XL-MIMO communication systems, where the conformal modular XL-ULA deployed at the BS is adopted to serve multiple single-antenna users. By exploiting the unique modular array architecture and considering the potential near-field propagation, the simplified USW models were first developed. Then, the beam focusing patterns were analyzed by taking into account near-field beamforming to realize near-field location focusing. It was revealed that modular XL-ULA can provide higher spatial resolution than its collocated counterpart, but at the cost of producing undesired grating lobes. Moreover, it was found that the SINR for users may be degraded severely by the grating lobes of the beam focusing pattern. To address this peculiar issue for modular architecture, an efficient greedy based user grouping method was proposed for multi-user transmission scheduling to maximize the users' sum rate. Numerical results verified the effectiveness of the proposed user grouping method, and the superiority of modular XL-MIMO over the collocated XL-MIMO for enhancing the system spectral efficiency, especially when there are densely distributed users.

\appendices

\section{Proof of Theorem 6}

To obtain the closed-form expression for \eqref{EQU-7231}, by setting $\left|\frac{F(x_{-3 \rm dB})}{x_{-3 \rm dB}}\right|= \frac{1}{2}$,  $x_{-3  \rm dB}\approx \pm 1.95$ is obtained. Therefore,
if $E_{\rm NF, NF}(z;\theta')$ is no less than $\frac{1}{2}$,  we have
\begin{equation}\label{EQU-223q} \vspace{-3pt}
\begin{split}
\left|z\right|\le\frac{4.84\lambda}{N^2\Gamma^2d^2\cos^2\theta'}\stackrel{(a)}{\approx} \frac{1}{r_{\rm hp}(\theta')},\\
\end{split}
\end{equation}
where $(a)$  holds due to $D\approx \Gamma Nd$ for large $N$. Based on \eqref{EQU-223q}, the effective distance resolution of modular XL-ULA under near-field beamforming with respect to $z$ is obtained as (24).
 Theorem 6 is therefore proved.\par
\section{Proof of Theorem 7}
Similar to Theorem 6, to obtain the closed-form expressions for (26) and (27), respectively, we need to consider $G_{\rm NF,NF}(r'+\Delta_r,\theta';r',\theta')$ is no less than $\frac{1}{2}$, thus leading to
\begin{equation}\label{EQU-223d} \vspace{-3pt}
\begin{split}
\left|\frac{\Delta_r }{\Delta_r r'+r'^2}\right|\le \frac{1}{r_{\rm hp}(\theta')}.\\
\end{split}
\end{equation}\par
To further simplify \eqref{EQU-223d}, we need to discuss the following two cases.\par
1) When $r_{\rm hp}(\theta')>r'$, \eqref{EQU-223d} is written as
\begin{equation}\label{EQU-22218} \vspace{-3pt}
-\frac{r'^2}{r_{\rm hp}(\theta')+r'}\le \Delta_r \le \frac{r'^2}{r_{\rm hp}(\theta')-r'}.
\end{equation}\par
2) When $r_{\rm hp}(\theta')\le r'$, \eqref{EQU-223d} is expressed as
\begin{equation}\label{EQU-22219} \vspace{-3pt}
-\frac{r'^2}{r_{\rm hp}(\theta')+r'}\le \Delta_r \le +\infty.
\end{equation}\par
Thus, the effective distance resolution of modular XL-ULA under  near-field beamforming in terms of $\Delta_r$ is obtained as (28) and (29).
The proof of Theorem 7 is thus completed.

\bibliographystyle{IEEEtran}
\end{document}